\documentclass[%
 reprint,
 amsmath,amssymb,
 aps,
pra,
]{revtex4-2}

\usepackage{graphicx}
\usepackage{dcolumn}
\usepackage{bm}
\usepackage[table]{xcolor}
\usepackage{amssymb}
\usepackage{amsmath}
\usepackage{xfrac}
\usepackage{relsize}
\usepackage{physics}

\begin{document}

\title{Pattern Forming Mechanisms of Color Vision}

\author{Zily Burstein}
 \email{Corresponding Author: ceburst@gmail.com}
\author{David D. Reid}%
\affiliation{%
 Department of Physics, University of Chicago, Chicago, IL, USA
}%

\author{Peter J. Thomas}
\affiliation{
 Department of Mathematics, Applied Mathematics, and Statistics; Department of Biology; Department of Cognitive Science, Case Western Reserve University, Case Western Reserve University, Cleveland, OH, USA
}%

\author{Jack D. Cowan}
\affiliation{%
 Department of Mathematics, University of Chicago, Chicago, IL, USA
}%

\begin{abstract}
While our understanding of the way single neurons process 
chromatic stimuli in the early visual pathway has advanced significantly in recent years, we do not yet know how these cells interact to form stable representations of hue. 
Drawing on physiological studies, we offer a dynamical model of how the primary visual cortex tunes for color, hinged on intracortical interactions and emergent network effects. 
After detailing the evolution of network activity through analytical and numerical approaches, we discuss the effects of the model's cortical parameters on the selectivity of the tuning curves. 
In particular, we explore the role of the model's thresholding nonlinearity in enhancing hue selectivity by expanding the region of stability, allowing for the precise encoding of chromatic stimuli in early vision. Finally, in the absence of a stimulus, the model is capable of explaining hallucinatory color perception via a Turing-like mechanism of biological pattern formation. 
\end{abstract}

\maketitle


\section{Introduction}\label{sec1}
Our experience of color begins in the early visual pathway, where, from the moment light strikes the retina, cone-specific neuronal responses set off the mechanisms by which the photons' chromatic information is converted to the hues we ultimately see. While color vision scientists agree that the single-cell processing of chromatic stimuli occurs along the two independent cone-opponent L$-$M and S$-$(L+M) pathways \citep{Conway, Boynton96}, there is yet no consensus as to how and where the divergent signals are synthesized to encode hue. To complicate matters, cone-opponency, observed in electrophysiological recordings of single neurons \citep{Shapley}, is often confounded with hue-opponency, a phenomenon first theorized by Ewald Hering in the nineteenth century and later mapped out in clinical studies by Jameson and Hurvich \citep{Shevell, sometransformations, Hurvich}.

\begin{figure}[h]
\centering
\includegraphics[width=0.4\textwidth]{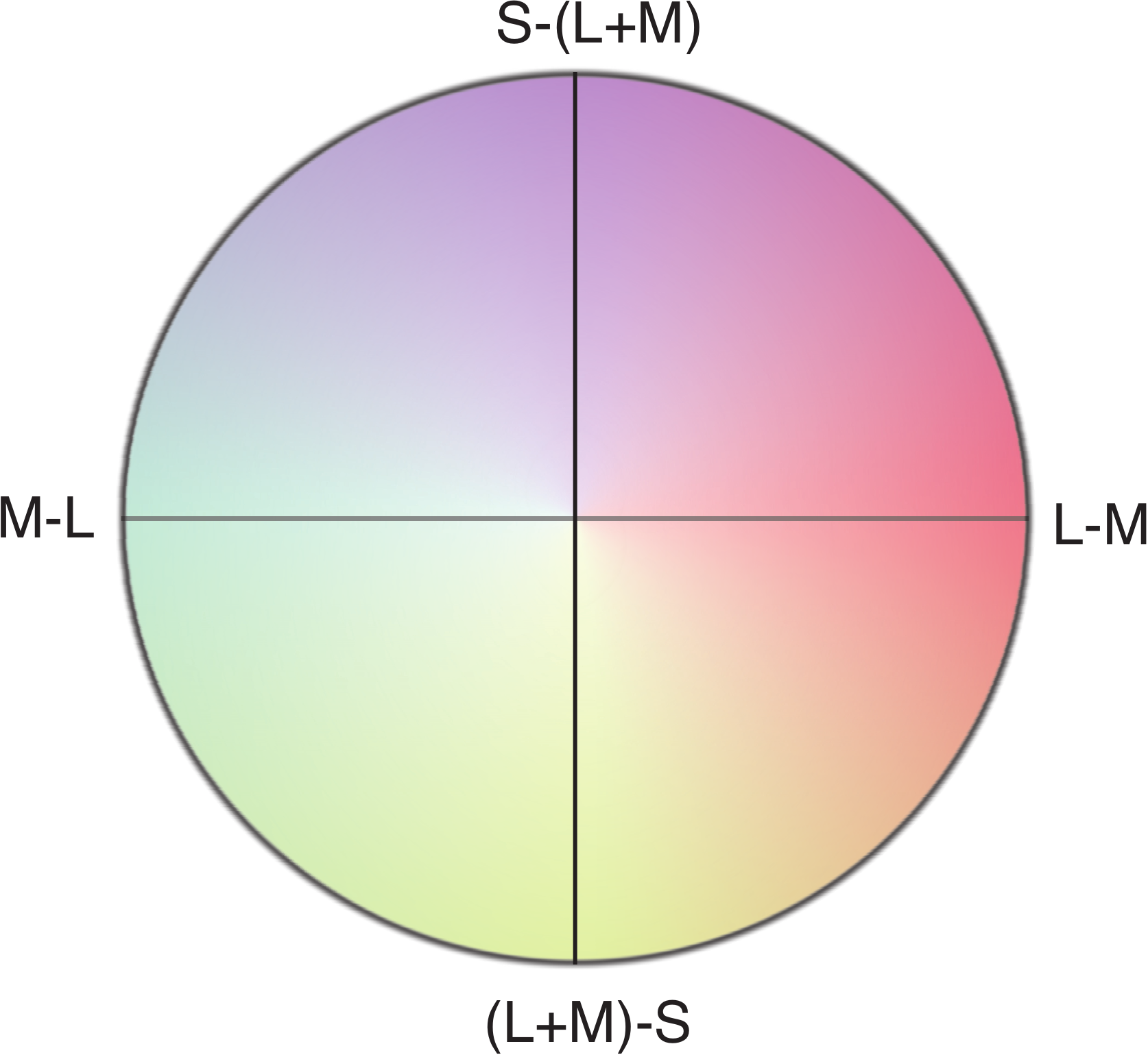}
\caption{The DKL space maps chromatic stimuli onto a circle with two ``cardinal'' axes representing the L$-$M and S$-$(L+M) pathways. The excitatory or inhibitory effect of a stimulus on cone-opponent cells tuned to either  pathway can be thought of as a projection of its location in DKL space onto the relevant axis. Stimuli isolating the two pathways correspond roughly to wavelengths associated with the red, green, blue, and yellow unique hues of color-opponent theory, leading to the common, but mistaken, conflation of chromatic stimulus and color perception.}
\label{fig. 1}
\end{figure}

Best depicted in the Derrington-Krauskopf-Lennie (DKL) stimulus space (Fig. \ref{fig. 1}), cone-opponency predicts that neurons tuned to either the L$-$M or S$-$(L+M) pathway will not respond to light whose wavelengths isolate the other \citep{DKL}. It is tempting to equate these null responses to the four unique hues of color-opponent theory, in which unique blue, for example, is observed when the ``redness'' and ``greenness'' of a perceived color exactly cancel. But the wavelengths of the unique hues specified by perceptual studies \citep{Hurvich} only roughly match the wavelengths isolating either cone-opponent pathway \citep{dklVhue, Wuerger, XiaoS}, and, more fundamentally, we do not yet understand the mechanisms behind the processing which the analogy implies \citep{Mollon, Stoughton, Valberg}. That is, how do we get from the single neurons' chromatic responses to our perception of color? 

The necessary processing has often been attributed to higher-level brain function \citep{Zaidi, Mehrani2020, Valois, Li, Lennie} or yet unidentified higher-order mechanisms \citep{Valberg, Wuerger}. A central question of color vision research is whether these mechanisms rely on parallel or modular processing to encode stimulus chromaticity \citep{Liu, Garg, Nauhaus, Shapley, Schluppeck, Conwaycolorarea}. If signalling about chromaticity is transmitted with information about other visual features, such as brightness, orientation, and spatial frequency, how do these features get teased apart? If not, where is the purported color center of the brain \citep{Gegenfurtner, Conwaymodules}?

Several authors have addressed these questions through combinatorial models which parameterize the weights of the L, M, and S cones contributing to successive stages of processing \citep{Stockman, Judd, Valois, Mehrani2020, Gegenfurtnerbook}. Though differing in their assumptions of modularity, the theories share a mechanistic framework for the transition of single-cell receptive field properties \citep{Tale}. Starting with cells in the retina and lateral geniculate nucleus (LGN) known to be tuned broadly to the cone-opponent axes, these proposed mechanisms build up to cells in various cortical areas more narrowly tuned to divergent (and debated) chromatic directions in DKL space.
While parsimonious, this formalism comes at the cost of tuning the cone weights arbitrarily, disregarding specific properties of real neurons' receptive fields \citep{Stockman, Eskew, Boynton96}. 
Furthermore, the linear combinatorial mechanism is not, on its own, able to account for the variety of color cells observed in the visual cortex \citep{Johnson, Garg, Shapley}. In addition to the forward flow of chromatic information through the successive stages of processing, the encoding of color reflects the neuronal dynamics within each.  
Modelers agree that the next forays into a mechanistic theory of color vision should consider these intracortical circuits, but disagree about where such interactions first become important \citep{Hanazawa, Valois, Liu, Wachtler}. 

Electrophysiological studies of macaque visual cortex have shed some light on this question, showing that the processing of individual hues previously associated with higher-level mechanisms has its origins in the primary visual cortex (V1) \citep{huemaps, cofd, Wachtler, Gegenfurtner, XiaoS, Garg, Hanazawa}.  These experiments have identified the emergence of neurons in V1 tuned to the gamut of hues in DKL space, as well as to the role of processing nonlinearities in determining their tuning curves \citep{sometransformations, Hanazawa, Wachtler, Lennie}. Puzzlingly, these cells mainly inhabit the so-called CO ``blobs,'' patchy regions rich in cytochrome oxidase that display a sensitivity to stimuli modulating either of the cone-opponent axes rather than the full set of hues \citep{LandismanCO, cofd, Livingston, Salzmann}. Some have speculated that this colocalization stems from a mixing of cell populations encoding the two cardinal pathways \citep{ cofd, XiaoS} while others indicate a distinct population of hue-sensitive neurons in the ``interblob'' regions, more conclusively associated with orientation tuning \citep{LandismanElectro, Garg}. As a whole, however, these studies point to the need for a population theory of chromatic processing remarkably early in the visual pathway.

In this article, we present a model of color processing in which intracortical neuronal dynamics within V1 serve as the substrate for hue perception. Drawing on the canonical Wilson-Cowan neural field equations and the ring model of orientation tuning, we show that this population approach allows us to account for cells responsive  to  the  full range  of DKL  directions without  the  need  to fine-tune input  parameters \citep{wilsoncowan1972, wilsoncowan1973, Ben-Yishai, Hansel, Zily}. The threshholding we employ bears in mind the input-response nonlinearities  of previous  combinatorial models,  but zooms out of the single-cell, feedforward interpretation of input as the stimulus-driven LGN afferents to individual neurons. Rather, we model input as the total synaptic current into a population of cells, taking into account both the cone-opponent LGN afferents as well as the hue-dependent connectivity between distinct neuronal populations.

The resulting demarcation between the cone-opponent and the hue-selective mechanisms in the same population of cells points to the importance of V1 in the transition from chromatic stimulus to color perception. To characterize this role, we study the effects of the model's connectivity parameters and processing nonlinearities on the narrowness and stability of the hue tuning curves. In the final part of the paper, we show that the model is able to explain color responses in the absence of LGN input, evoking color hallucinations via a Turing-like mechanism of spontaneous pattern formation in DKL space. 

\section{Model}\label{sec2}
In light of the patchy distribution of color-sensitive cells reported in \cite{Livingston, cofd, LandismanCO}; and \cite{Salzmann}, we model the color map of V1 as a set of neuronal networks, each encoding the chromaticity of its corresponding region of the visual field. This organization brings to mind the hypercolumnar structure of orientation preference within V1 \citep{Hubel}, which, on the basis of its feature-based connectivity properties, allows for the representation of network activity as a function of a localized feature space. Here, we assume a mean hue-dependent activity $a(\theta,t)$ where $\theta$ represents a direction in the DKL stimulus space, a strictly physiological conception of ``hue'' from the hues categorizing color perception, as explained above. In drawing this distinction, and in agreement with \cite{dklVhue} and \cite{cofd}, we give no special status to V1 cells tuned to the DKL directions associated with the unique hues of color-opponent theory, while simultaneously emphasizing the cone-opponent nature of feedforward afferents from the LGN. 

The resulting activity $a(\theta,t)$ of a network of hue-preferring cells, expressed as a firing rate in units of spikes/second, is dominated by the membrane properties of its constituent cells, whose potential variations occur on the order of the membrane time constant $\tau_{0}$, taken to be 10 msec \citep{Izhikevich, Ben-Yishai, Carandini}. In the vein of previous neural mean-field models of feature detection \citep{Crystal, Spherical, Geometric, Dayan, ErmentroutReview,Ermentrout2003}, and in close analogy to the ring model of orientation tuning \citep{Hansel, Ben-Yishai}, we let $a(\theta,t)$ evolve according to the single-population firing-rate formulation of the Wilson-Cowan equations:
\begin{equation}
\label{WC}
    \tau_{0}\frac{da(\theta, t)}{dt} =-a(\theta, t) + g[h(\theta,t)],
\end{equation}
where $h(\theta,t)$, the synaptic input, takes into account both excitatory and inhibitory afferents into a population of cells preferring hue $\theta$, and $g(h)$ is an activation function, as described below. 

To analyze the relationships between feedforward and recurrent processing and to distinguish between their respective effects on $a(\theta,t)$, we write $h(\theta,t)$ as a sum of the stimulus-driven synaptic input from the LGN and the intracortical input resulting from the hue-dependent network connectivity within V1: 
\begin{equation}
\label{total input}
h(\theta,t)=h_\text{ctx}(\theta,t)+h_\text{lgn}(\theta).
\end{equation}

We express the input as the average effect of the net synaptic current on the membrane potential of a cell, following the conventions of \cite{ErmentroutReview} and \cite{Carandini}. Thus, $h(\theta,t)$ has units of mV and can take on both positive and negative values, chosen here so that $a(\theta,t)$ typically ranges from 0 to 60 spikes/second, consistent with electrophysiological experiments penetrating individual color-responsive cells \citep{Wachtler, Johnson, LandismanElectro, conway2001spatial}.

The input is converted to a firing rate according to the nonlinear activation function
\begin{equation}
\label{activation}
g(h)=\beta\cdot(h-T) \cdot \mathcal{H}(h-T),
\end{equation}
where $\mathcal{H}(x)$ is the Heaviside step function defined as $\mathcal{H}(x)=1$ for $x>0$ and zero for $x\leq0$. Note that in the context of machine learning, this form of activation is also known as the rectified linear unit function, or ReLU for short. By constraining the network activity to levels below 60 spikes/second, we ignore the effects of neuronal saturation commonly implemented in models of $g(h)$ \citep{ErmentroutReview, Ben-Yishai}. Here, $T$ is the threshold potential of a neuron, below which the synaptic input has no effect on the mean firing rate of the network. Interestingly, as a processing feature, this thresholding nonlinearity has been speculated to account for the chromatic responses of individual neurons in V1 \citep{Hanazawa}. The amplification of these responses, and thus the mean network response, is modulated by $\beta$, the neural gain measured in spikes$\cdot$sec$^{-1}$/mV. We assume that $\beta$ is determined by far-ranging internal and external influences, from attentional mechanisms to hallucinogenic input \citep{gain, LSD}.

\subsection{Feedforward Input}\label{subsec2}
To parameterize the input, prior work has relied on the direct relationship between cortical feature preferences and properties of the visual stimulus \citep{Spherical, Ben-Yishai}. Cells in the cortex labeled, for instance, by their spatial frequency preferences can be mapped directly onto a visual space parameterized by the same variable. Thus, the activity of each neuronal population is no longer labeled purely by its position on the cortical sheet, but also by its preferred stimulus in an analogous feature space. 

The corresponding network topology may be  modeled on the cortical histology, such as the orientation map of \cite{Bosking} or spatial-frequency maps addressed in \cite{Crystal}, \cite{Horizontal}, and \cite{Spherical}. Conversely, it may be based entirely on functional considerations, as for instance in the orientation tuning model of Sompolinksy et al., also known as the ``ring model,'' which posits a topology based on the experimentally-motivated assumption that populations with similar orientation preferences are maximally connected \citep{Ben-Yishai} and on the argument that the important features of such a connectivity are captured by its first-order Fourier components \citep{Hansel}. 

Our model deviates in this regard by emphasizing that the stimulus's chromatic information is first discretized along the two cone-opponent pathways. We incorporate this aspect of early processing by projecting the stimulus's DKL space position $\bar{\theta}$ onto the two cardinal axes:
\begin{align}
l&=\cos\bar{\theta} \nonumber \\
s&=\sin\bar{\theta}.
\end{align}
The magnitudes of $l$ and $s$ are thus taken to represent the normalized strengths of the L$-$M and S$-$(L+M) cone-opponent signals respectively. The feedforward input is then given by
\begin{equation}
    \label{lgn_input1}
    h_\text{lgn}=c(l\cos\theta+s\sin\theta),
\end{equation}
where $c$ is the signal strength, or contrast, expressed as the mean postsynaptic \emph{coarse} membrane potential (in units of mV) of the target hue population generated by the presynaptic LGN neurons \citep{CarandiniCat}. Formulated in this way, the input captures the colocalization of cone-opponency and hue selectivity in the activity of V1 cells as observed in \cite{cofd} and \cite{huemaps}. The hue-tuning networks, parameterized by $\theta$, are not only responsive to the individual cone-opponent stimulus signals, $l$ and $s$, but also implement the combinatorial mechanisms by which they are first mixed \citep{sometransformations}. Substituting the expressions for $l$ and $s$ into \ref{lgn_input1}, we obtain
\begin{equation}
\label{lgn}
h_\text{lgn}=c\cos(\theta-\bar{\theta}).
\end{equation}
With this form, we point out the similarity of our combinatorial scheme to that of \cite{Mehrani2020}, in which the input from cone-opponent V2 cells into hue-tuning V4 cells is weighted as a function of the difference in their preferred hue angles. Most evidently, we differ from this model by first combining the cone-opponent signals in V1 rather than V4, in accordance with the above-mentioned studies. But beyond pointing to V1 as the origin of mixing, these experiments indicate that the combinatorial feedforward scheme is not sufficient to account for the variability of neuronal hue preferences. \cite{cofd} showed, for instance, that the contribution of signals isolating the S$-$(L+M) pathway is too small to explain the shifting of hue preferences away from the L$-$M axis by purely combinatorial means. As put forward by \cite{Shapley}, \cite{Wachtler}, and \cite{Lehky}, a more complete understanding of neuronal hue encoding within V1 requires us to consider the nonlinear population dynamics therein.

\subsection{Recurrent Interactions}
We begin by characterizing the connectivity of the target hue tuning populations with a translation invariant cortical connectivity function $w(\lvert x-x'\rvert)$, such that the interactions between neurons in a single CO blob (length scale $\sim$ 0.5 mm) depend only on the cortical distance between them \citep{distancedependence, Salzmann}. The network's connectivity comprises the interactions of both its excitatory and inhibitory populations,
\begin{equation}
 w(\lvert x-x'\rvert)=w_\text{exc}+w_\text{inh},
\end{equation}
each of which we model as a sum of an isotropic and distance-dependent term: 
\begin{align}
    \label{exc_inh}
    w_\text{exc}&=E_{0}+E_{1}\cos(\lvert x-x'\rvert) \nonumber
   \\ w_\text{inh}&=-I_{0}-I_{1}\cos(\lvert x-x'\rvert).
\end{align}
We set $E_{0}{\geq}E_{1}{>}0$ and $I_{0}{\geq}I_{1}{>}0$ so that $w_\text{exc}$ and $w_\text{inh}$ are purely excitatory and inhibitory, respectively, in accordance with Dale's Law \citep{Dayan, Ben-Yishai}.

Next, we map the weighting function onto hue space, drawing from the hue-tuning micro-architecture revealed by the imaging studies of \cite{Liu} and \cite{huemaps}. These studies point to a linear relationship between distance and hue angle difference, which minimizes the wiring length of cells tuned to similar hues \citep{Liu}. The hue-preferring cells inhabit the so-called ``color regions,'' defined as such for their activation by red-green grating stimuli \citep{Liu}. These regions predominantly overlap with the V1 CO blobs \citep{cofd, LandismanCO} and are responsive to the full range of hues, much like the patchy distribution of orientation maps within the V1 hypercolumns. Thus, in a similar manner to the local feature processing models of \cite{Spherical} and \cite{Ben-Yishai}, we model the CO blob as a single color-processing unit consisting of $N$ neurons labeled by the continuous hue preference variable $\theta \in [-\pi, \pi]$ \citep{Spherical}.

\begin{figure}[h]
\centering
\includegraphics[width=0.85\textwidth]{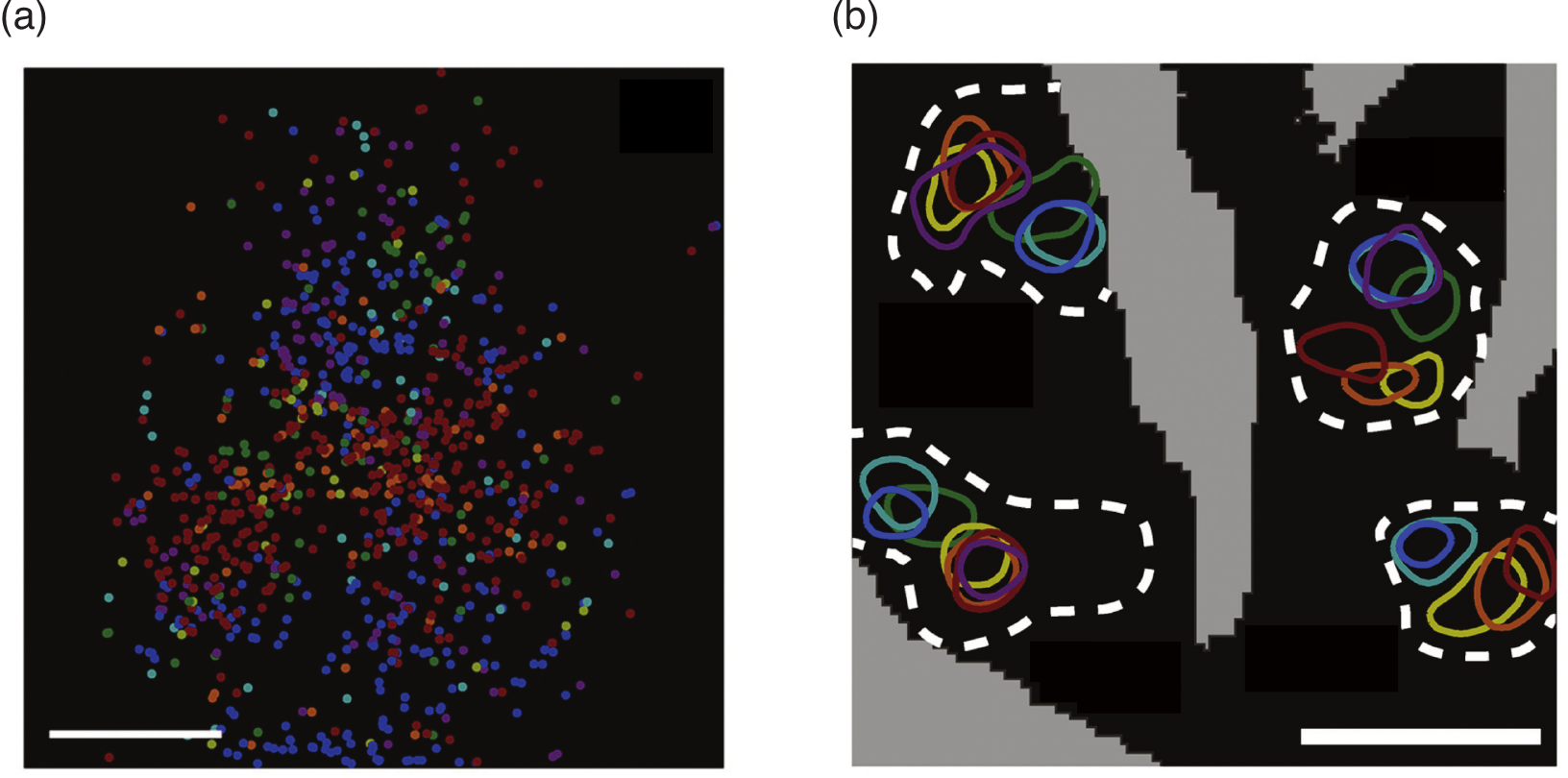}
\caption{(a) Hue map of individual hue-selective cells obtained by 2-photon calcium imaging of neuronal responsiveness to seven test hues. Scale bar: 200 \textmu m. (b) Regions of peak response to test hues (solid contours). The dashed white lines demarcate the color-preferring regions, colocalized with the CO blobs. Scale bar: 0.5 mm. Modified with permission from \cite{Liu}.}
\label{fig. 2}
\end{figure}

Figure \ref{fig. 2} shows the distribution of hue-responsive neurons within a typical color region (Fig. \ref{fig. 2}a) as well as a more coarse-grained demarcation of peak activity within several of these regions (Fig. \ref{fig. 2}b). To describe the spatial organization of their hue preference data, \cite{huemaps} and \cite{Liu} applied a linear fit to the cortical distance between two cell populations as a function of the difference in their preferred hue stimuli $\Delta \theta \equiv \lvert \theta-\theta'\rvert$ apart in DKL space. Note, this implies a discontinuity between $\theta=0$ and $\theta=2 \pi$, allowing for the $2 \pi$ periodicity of the hue preference label. \cite{Liu} report that the linear fit was able to capture the micro-organization of $42\%$ of their tested hue maps, and a regression performed by \cite{huemaps} on an individual hue map gave a squared correlation coefficient of $R^{2}=0.91$. 

\onecolumngrid

\twocolumngrid

\begin{figure}[h]
\centering
\includegraphics[width=0.4\textwidth]{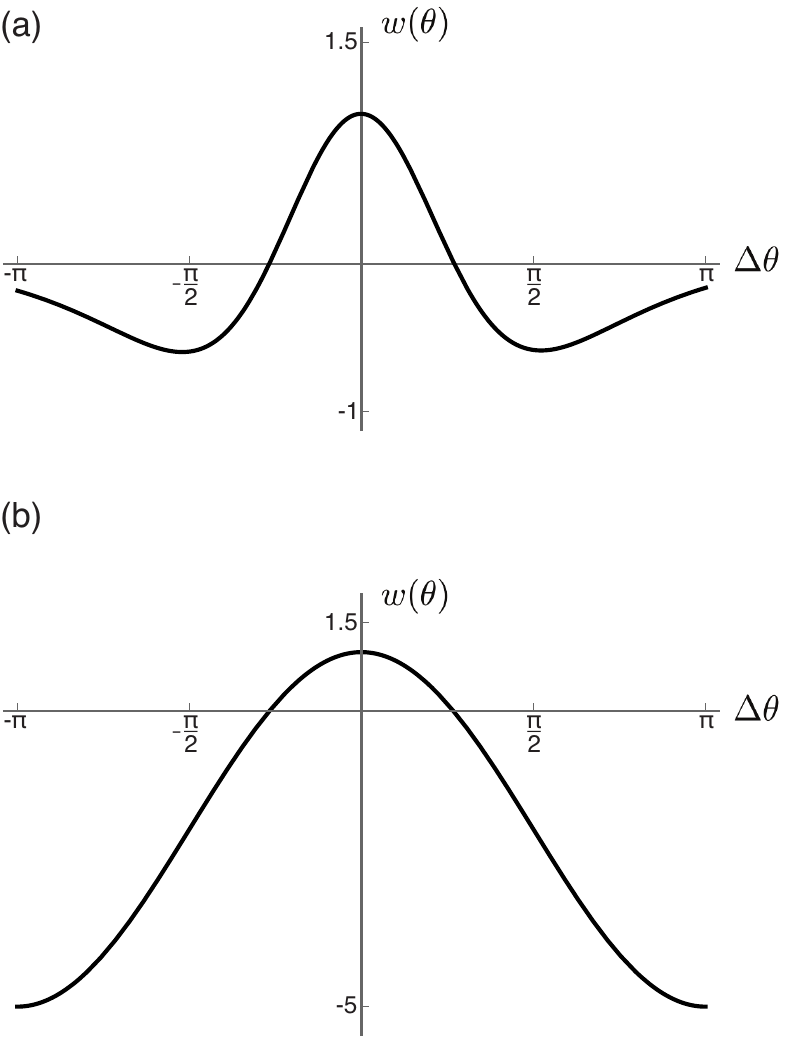}
\caption{Cortical connectivity functions. (a) A difference of two Gaussians, one characterizing the excitatory interactions (here with $\sigma_{E} = 40 ^{\circ} $) and the other the inhibitory interactions ($\sigma_{I} = 90 ^{\circ}) $. This is the connectivity typically assumed in mean field models of cortical processing. (b) The difference of cosines formulation (\ref{differenceofcosines}), with $J_{0}=-2$ and $J_{1}=3$, captures the local excitation and lateral inhibition assumed in (a). }
\label{weighting}
\end{figure}

In agreement with these findings, we let $\lvert x-x'\rvert = \lvert \theta-\theta' \rvert $, absorbing the regression parameters into the connectivity strength values $E_{0}$, $E_{1}$, $I_{0}$, and $I_{1}$ in \ref{exc_inh}. Substituting this change of variables and setting $J_{0}=E_{0}-I_{0}$, $J_{1}=E_{1}-I_{1}$ (measured in mV/spikes$\cdot$sec$^{-1}$) gives
\begin{equation}
    \label{differenceofcosines}
    w(\theta-\theta')=J_{0}+J_{1}\cos(\theta-\theta').
\end{equation} 
As detailed in Fig. \ref{weighting}, for $J_{1}>0$, this functional form captures the local excitation and lateral inhibition connectivity ansatz typically assumed in neural field models as an analogy to diffusion-driven pattern formation \citep{Amari, Hoyle, Turing, Kim, BressloffTuring}. Notably, neurons in close proximity in both cortical and hue space maximally excite each other, and those separated by $\Delta \theta=\pi$ maximally inhibit each other, evoking the hue-opponency of perception on a cellular level. We emphasize, however, that this choice of metric is guided by our physiological definition of hue and does not associate a perceived color difference to measurements in hue space.

Here, it is also important to distinguish between the connectivity function and the center-surround receptive fields of single- and double-opponent color cells \citep{Shapley}. While the structures of both can be approximated by the same functional form, the resemblance is superficial: the former characterizes the interactions between different neuronal populations, and the latter is a property of single cells, often adapted for computer vision algorithms \citep{Somers, Turner}. 
 
Finally, we weigh the influence of the presynaptic cells by convolving the connectivity function with the network activity, arriving at the cortical input to the target hue population at time $t$: 
\begin{equation}
\label{corticalinput}
h_\text{ctx}(\theta,t)=\int_{-\pi}^\pi w(\theta-\theta')a(\theta',t)d\theta'.
\end{equation}
The recurrent input is thus a continuous function in $\theta$, derived from the population-level interactions. As put forward by the above-mentioned imaging studies, these interactions are colocalized with the cone-opponent feedforward input, $h_\text{lgn}$, within the same CO blob regions of V1. Collectively, our formulation of $h(\theta,t)$ implements the mixing rules posited by these experiments, without requiring us to arbitrarily fine-tune the relative weights of the afferent signals.

\section{Results}\label{sec3}

\subsection{Evolution of Network Activity}
We start by observing that by virtue of the invariance of $w(\theta-\theta')$ under translations of $\theta$, the convolution operator $T_{w} \rightarrow w*f(\theta)=\int_{-\pi}^\pi w(\theta-\theta')f(\theta')d\theta'$ is diagonalizable by the Fourier eigenfunction basis
\begin{equation}
    \label{basis}
   {\hat{e}_{\mu}}(\theta)=\frac{1}{\sqrt{2\pi}}e^{i \mu \theta}
\end{equation}
with $\mu \in \mathbb{N}$ and $ {\hat{e}_{\mu}}$ normalized to integrate to $1$ on $[-\pi,\pi]$. To calculate the eigenvalues $\lambda _{\mu}$ of the corresponding linear transformations,
\begin{equation}
    \label{linear transformation}
    \int_{-\pi}^\pi w(\theta-\theta')\frac{1}{\sqrt{2\pi}}e^{i \mu \theta'}d\theta'=\lambda _{\mu}\frac{1}{\sqrt{2\pi}}e^{i \mu \theta},
\end{equation}
we make the change of variables $\theta-\theta'=\phi$, so that the left-hand side of \ref{linear transformation} can be rewritten as
\begin{flalign}
    -\int\limits_{\theta+\pi}^{\theta-\pi}\frac{w(\phi)e^{-i \mu \phi}e^{i \mu \theta}}{\sqrt{2\pi}} d\phi =
    \int\limits_{-\pi}^\pi \frac{w(\phi)e^{-i \mu \phi}e^{i \mu \theta}}{\sqrt{2\pi}} d\phi.
\end{flalign}

The eigenvalues are thus:
\begin{equation}
    \lambda _{\mu}=\int_{-\pi}^\pi w(\phi)e^{-i \mu \phi} d\phi.
\end{equation}

Next, we assume $a(\theta,t)$ is separable in $t$ and $\theta$ and bounded on $[-\pi,\pi]$ so that we may expand it in the eigenbasis of the convolution operator as: 
\begin{equation}
\label{decompose}
a(\theta,t)=\sum \limits_{\mu} c_{\mu}(t){\hat{e}_{\mu}(\theta)}.
\end{equation}
Substituting the expansion into \ref{corticalinput}, we have for \ref{total input}
\begin{equation}
    h(\theta,t)=\sum  \limits_{\mu} \left [ c_{\mu}(t)\int\limits_{-\pi}^\pi w( \theta-\theta'){\hat{e}_{\mu}}(\theta')d\theta'\right]+h_\text{lgn}(\theta),
\end{equation}
where $w( \theta-\theta')$  is our choice for the connectivity function (\ref{differenceofcosines}) and $h_\text{lgn}(\theta)$ is defined as in \ref{lgn}. Evaluating the integrals, we obtain
\begin{align}
 \begin{split}
    \label{h_lambda}
    h(\theta,t)=\lambda_{\text{-} 1} \, c_{\text{-} 1}(t)\,{\hat{e}_{\text{-} 1}}(\theta)+\lambda_{0}\,c_{0}(t)\,{\hat{e}_{0}}(\theta) \\
    +\lambda_{1}\,c_{1}(t)\,{\hat{e}_{1}}(\theta)+c\cos(\theta-\bar{\theta}),
 \end{split}
\end{align}
with $\lambda_{0}=2 \pi J_{0}$ and $\lambda_{1}=\lambda_{\text{-} 1}=\pi J_{1}.$ Note here that only the zeroth and first-order complex Fourier components remain.

Substituting the expansion \ref{decompose} and the explicit form of the activation function \ref{activation} into \ref{WC} yields:
\begin{flalign}
\label{ratedecomposed}
\tau_{0}\sum \limits_{\mu = -\infty}^{\infty} \frac{dc_{\mu}(t)}{dt}{\hat{e}_{\mu}(\theta)} = -\sum \limits_{\mu =-\infty}^{\infty}c_{\mu}(t)\hat{e}_{\mu}(\theta) \nonumber \\
 +\beta \big(h(\theta,t)-T\big)\mathcal{H}\big(h(\theta,t)-T\big).
\end{flalign}

In the absence of the nonlinearity, each of the eigenmodes ${\hat{e}_{\mu}(\theta)}$ would evolve independently of the others, and a complete analysis of the time-dependent system would seek to solve a set of equations for $c_{\mu}(t)$ (see \emph{Appendix A: Linear Solution}). However, in our setup, the thresholding introduces a coupling of these coefficients, as the critical hue angles, $\delta_{1}$ and $\delta_{2}$, at which the input is cut off is determined by the combined $c_{\mu}(t)$ at each point in time. While an analytical solution to this system is in most cases intractable, it is nonetheless informative to break down the rate equation to a coupled system of equations for the evolution of the coefficients $c_{\mu}(t)$. Taking the inner product of \ref{ratedecomposed} with $\hat{e}_{\nu}$ and using $\bra{\hat{e}_{\nu}} \ket{\hat{e}_{\mu}}=\delta_{\mu \nu}$, we obtain:
\begin{align}
    \label{coefficientnet}
    \tau_{0}\frac{dc_{\nu}(t)}{dt}&=-c_{\nu}(t)+\bra{\hat{e}_{\nu}}\ket{\beta\big(h-T\big)\mathcal{H}\big(h-T\big)} 
    \nonumber \\ 
    &=-c_{\nu}(t)+\beta\int_{\delta_{1}(t)}^{\delta_{2}(t)}h(\phi,t)\hat{e}^{*}_{\nu}(\phi)d\phi
\end{align}
where the Heaviside restricts the domain of the inner product to $[\delta_{1}(t)$, $\delta_{2}(t)]$. The time dependence of the cutoff angles reflects the evolution of this curve, which requires that the thresholding be carried out continuously throughout the duration of the dynamics.

To determine $\delta_{1}$ and $\delta_{2}$, we reformulate the Heaviside as a function of $\theta$. Given that the input $h(\theta,t)$ is a real-valued function, $c_{0} \in \mathbb{R}$ and $c_{1}=c_{\text{-}1}^{*}$. For mathematical convenience, we then rewrite \ref{h_lambda} in terms of $c_{0}$, Re($c_{\text{-}1}$)$\equiv c_{\text{-}1}^{R}$, and Im($c_{\text{-}1}$)$\equiv c_{\text{-}1}^{I}$ as
\begin{align}
\begin{split}
    \label{h_rewritten}
h(\theta,t)=\frac{\lambda_{0} \, c_{0}(t)}{\sqrt{2\pi}}+\left(cl+\sqrt{\tfrac{2}{\pi}}\lambda_{\text{-}1}\,c_{\text{-}1}^{R}(t)\right)\cos(\theta) \\ +\left(cs+\sqrt{\tfrac{2}{\pi}}\lambda_{\text{-}1}\,c_{\text{-}1}^{I}(t)\right)\sin(\theta).
\end{split}
\end{align}
Setting 
\begin{align}
\label{subs}
q_\text{{R}}&=cl+\sqrt{\tfrac{2}{\pi}}\lambda_{\text{-}1}\,c_{\text{-}1}^{R}(t) \nonumber \\  q_\text{{I}}&=cs+\sqrt{\tfrac{2}{\pi}}\lambda_{\text{-}1}\,c_{\text{-}1}^{I}(t) \nonumber \\ 
q_{0}&=\frac{\lambda_{0} \, c_{0}(t)}{\sqrt{2\pi}}
\end{align}
the input takes the form
\begin{equation}
\label{h_v_theta}
    h(\theta,t)= q_{0}(t)+c_{h}(t)\cos[\theta +\gamma(t)]
\end{equation}
where $\tan(\gamma)=-\frac{q_\text{{I}}}{q_\text{{R}}}$ and $c_{h}(t)=\sqrt{\scriptstyle{q_\text{{R}}^{2}+q_\text{{I}}^{2}}}$.

\begin{figure}[ht]
\centering
\includegraphics[width=0.4\textwidth]{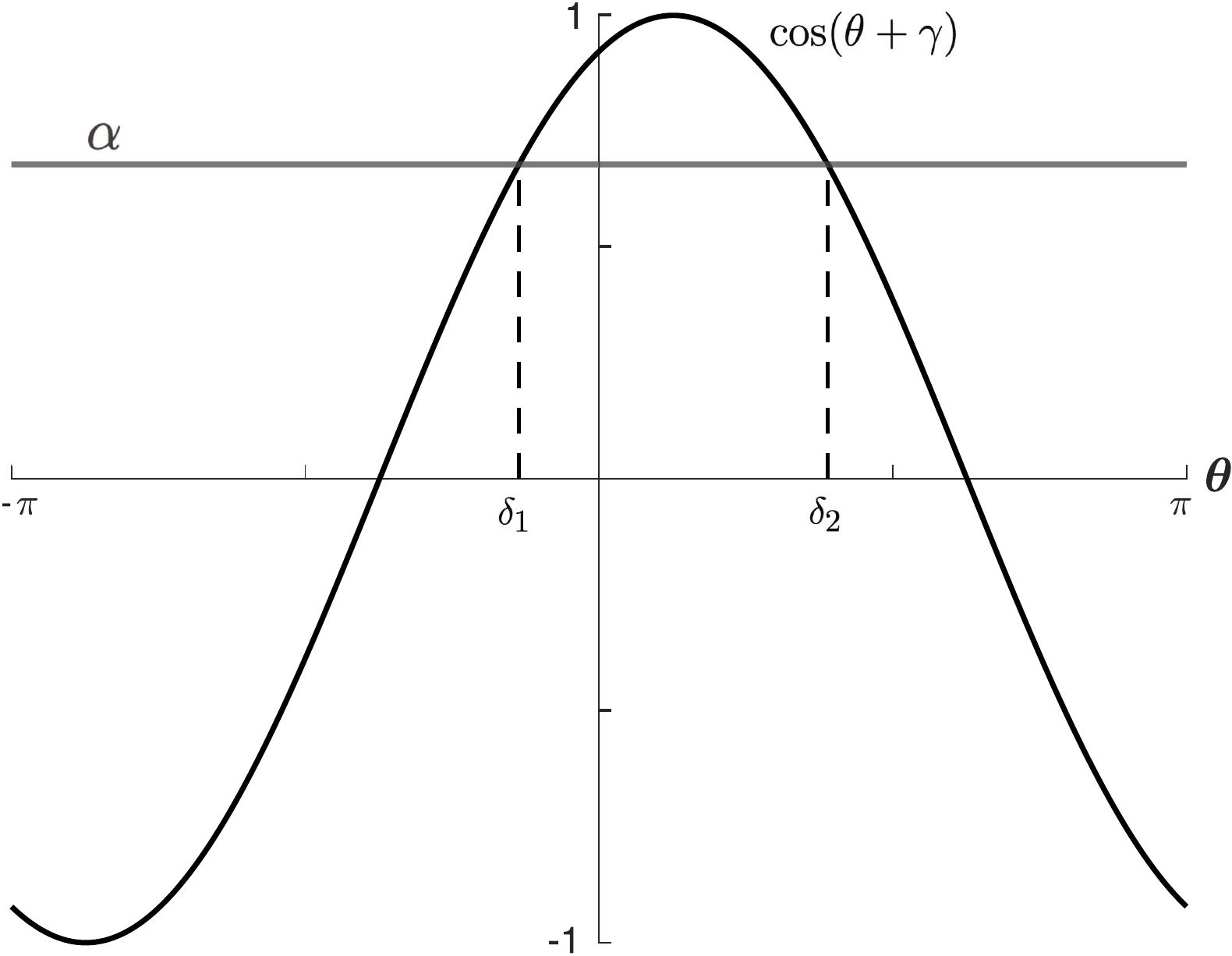}
\caption[Obtaining $\delta_{1}$ and  $\delta_{2}$]{The limits of integration $\delta_{1}$ and  $\delta_{2}$ in \ref{coefficientnet} are the angles corresponding to the intersection of $\alpha$ (in gray) and $\cos(\theta +\gamma)$ (in black). Here, $c=1$, $\beta=1$, and $T=-1$. $\bar{\theta}=\pi /8$. $J_{0}$ and $J_{1}$ are as in Fig. \ref{weighting}.}
\label{delta1delta2}
\end{figure}

The Heaviside can then be expressed as 
\begin{align}
    \label{Heaviformulation}
    \mathcal{H}\big[h-T\big]&=\mathcal{H}\big[q_{0}+c_{h}\cos(\theta +\gamma)-T\big] \nonumber \\
    &=\mathcal{H}\left[\cos(\theta +\gamma)-\alpha \right]
\end{align}
where $\alpha \equiv \tfrac{T-q_{0}}{c_{h}}$, and the time arguments are suppressed for simplicity. In this formulation, the Heaviside sets the limits of integration in \ref{coefficientnet} as the angles $\theta=\delta_{1}, \delta_{2}$ where $\alpha$ intersects with $\cos(\theta +\gamma)$, as shown in Fig. \ref{delta1delta2}. 

With this reformulation, the system of equations for the evolution of the coupled $c_{\nu}$ (\ref{coefficientnet}) takes the more explicit form:
\begin{align}
\begin{split}
\label{systemuneval}
&\tau_{0}\frac{dc_{\nu}(t)}{dt}=-c_{\nu}(t) \\
&+\beta\int_{\delta_{1}}^{\delta_{2}}\big[q_{0}(t)+c_{h}(t)\cos[\phi +\gamma(t)]\big]\hat{e}^{*}_{\nu}(\phi)d\phi. 
\end{split}
\end{align} 

Note that, for all $c_\nu$, the integrand of \ref{systemuneval} is a function of $q_{0}(t)$, $c_{h}(t)$, and $\gamma(t)$ and therefore, implicitly, only of the coefficients $c_{0}(t)$, $c_{\text{-} 1}(t)$, and $c_{1}(t)$. Thus, the dynamics are determined in full by the evolution of $c_{\lvert \nu \rvert \leq1}(t)$:
\begin{align}
\label{systemofequations}
    \tau_{0}\frac{dc_{0}(t)}{dt}&=-c_{0}(t)+\frac{\beta}{\sqrt{2\pi}}\int_{\delta_{1}}^{\delta_{2}}\left[h(\phi,t)\right]d\phi \nonumber \\
    \tau_{0}\frac{dc_{1}(t)}{dt}&=-c_{1}(t)+\frac{\beta}{\sqrt{2\pi}}\int_{\delta_{1}}^{\delta_{2}}\left[h(\phi,t)\right]e^{-i\theta}(\phi)d\phi \nonumber \\
    \tau_{0}\frac{dc_{\text{-}1}(t)}{dt}&=-c_{\text{-}1}(t)+\frac{\beta}{\sqrt{2\pi}}\int_{\delta_{1}}^{\delta_{2}}\left[h(\phi,t)\right]e^{i\theta}(\phi)d\phi, 
\end{align}
with $h(\phi,t)$ as in \ref{h_v_theta}.

Separating \ref{systemofequations} into its real and imaginary parts, and noting that a real-valued activity profile $a(\theta,t)$ requires $c_{0} \in \mathbb{R}$ and $c_{1}=-c_{\text{-1}}^{*}$, reduces the system to a set of equations for $c_{0}(t)$, $c_{\text{-}1}^{R}(t)$, and $c_{\text{-}1}^{I}(t)$.

Evaluating the integrals, we obtain:
\begin{widetext}
\begin{align}
\label{systemofequationseval}
\begin{split}
\tau_{0}\frac{dc_{0}(t)}{dt}&= -c_{0}(t)+\tfrac{\beta}{\sqrt{2\pi}}\Big\{ c_{h}\big[\sin(\delta_{2}+\gamma)-\sin(\delta_{1}+\gamma)\big]+(T-q_{0})(\delta_{1}-\delta_{2})\Big\} \nonumber
\end{split}
\\[2ex]
\begin{split}
\tau_{0}\frac{dc_{\text{-}1}^{R}(t)}{dt}&= -c_{\text{-}1}^{R}(t)+\tfrac{\beta}{\sqrt{2\pi}}\Big\{\tfrac{c_{h}}{2}\big[\cos{\gamma}(\delta_{2}-\delta_{1})+\cos(\gamma+\delta_{1}+\delta_{2})\sin(\delta_{2}-\delta_{1})\big] \\ &\qquad+(T-q_{0})(\sin\delta_{1}-\sin\delta_{2})\Big\} \nonumber
\end{split}
\\[2ex]
\tau_{0}\frac{dc_{\text{-}1}^{I}(t)}{dt}&= -c_{\text{-}1}^{I}(t)+\tfrac{\beta}{\sqrt{2\pi}}\Big\{\tfrac{c_{h}}{2}\big[\sin{\gamma}(\delta_{1}-\delta_{2})+\sin(\gamma+\delta_{1}+\delta_{2})\sin(\delta_{2}-\delta_{1})\big] 
\notag \\ &\qquad+(T-q_{0})(\cos\delta_{2}-\cos\delta_{1})\Big\}, 
\end{align}
\end{widetext}
where the time arguments of $q_{0},$ $c_{h},$ $\gamma,$ $\delta_{1},$ and $\delta_{2}$ are suppressed for clarity. 

Written in this form, the system provides a representation of the time evolution of $a(\theta,t)$ in terms of the coupled evolution of the constants $c_{\lvert \nu \rvert \leq1}$. It is important to note that these equations are nonlinear due to the implicit Heaviside in our determination of $\delta_{1}(t)$ and $\delta_{2}(t)$. While our reformulation of the right-hand side of \ref{systemuneval} allows for the explicit representation of the coupling of $c_{\nu}$ via the nonlinearity, it is also this coupling which proves the analytical solution of the trajectories intractable. Thus, to describe the behavior of the time-dependent solution, we turn next to a numerical analysis of the system's phase portrait --- that is, to an exploration of the features and stability of the system's emergent steady states.

\subsection{Steady-State Solution}
\label{numericalsolution}
We approach the solution to \ref{WC} with a Forward Euler method, propagating the activity from a random array of spontaneous initial values between $0$ and $0.2$ spikes/sec to its steady-state value. Within each timestep (typically chosen to be $1$ msec), we coarse-grain the network into $n=501$ populations with hue preferences separated evenly across the DKL angle domain $[-\pi,\pi]$. The choice of an odd $n$ allows us to numerically integrate \ref{corticalinput} using the Composite Simpson's Rule, whereupon we rectify $\{h(\theta,t)-T\}$ and evaluate the right-hand side of \ref{WC}. Below, we use the term \emph{tuning curve} only in reference to the emergent steady-state activity profiles. 

\begin{figure}[h]
\centering
\includegraphics[width=0.4\textwidth]{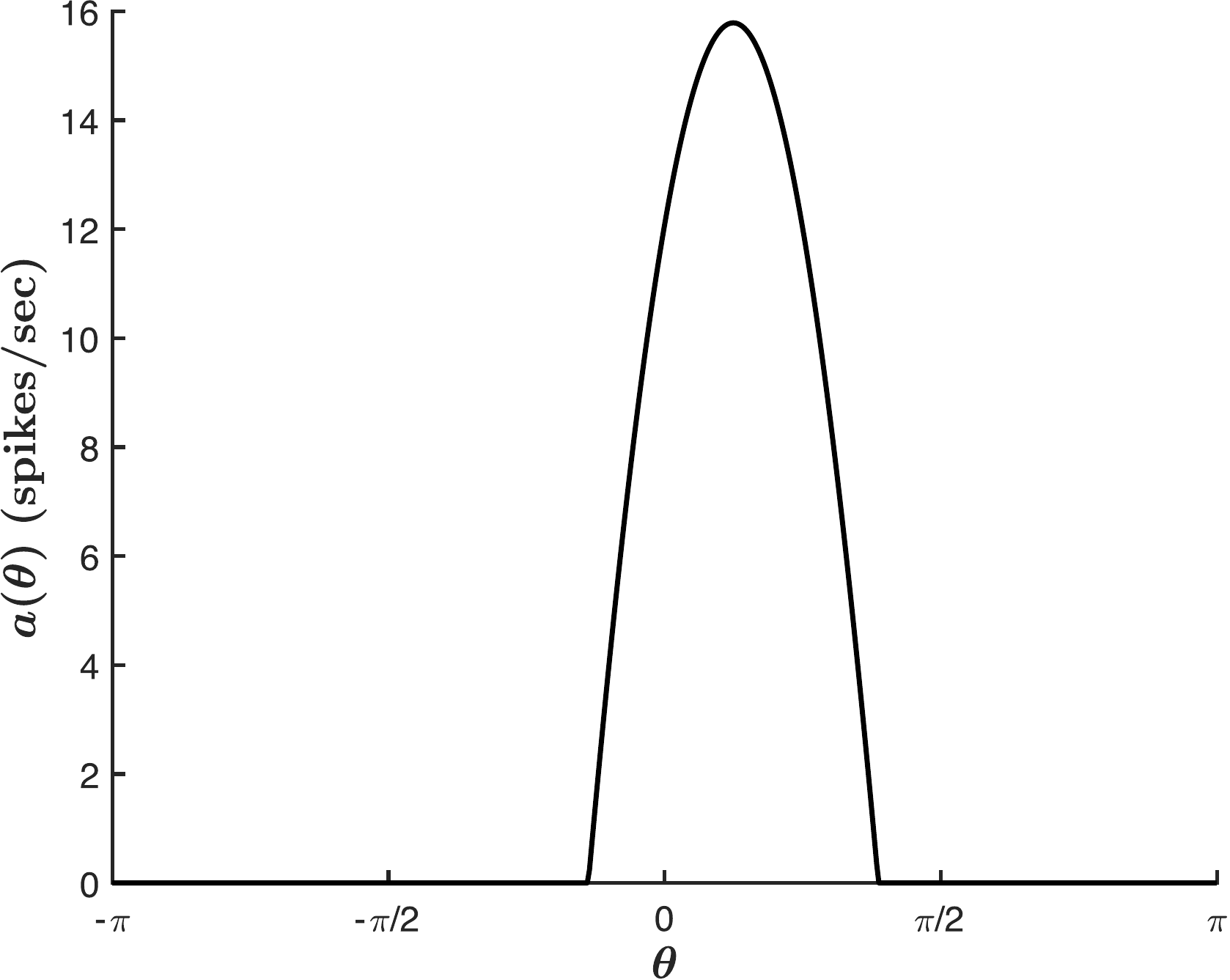}
\caption{Steady-state activity profile for a neuronal network encoding stimulus $\bar{\theta}=\pi/8$. Parameters are as in Fig. \ref{delta1delta2}.}
\label{generalcurve}
\end{figure}

Figure \ref{generalcurve} shows an example of a hue tuning curve obtained with this method. Note that the peak of the tuning curve is located at the LGN hue input angle $\bar{\theta}$, which is equivalent to the steady-state value of $-\gamma$ in \ref{systemofequationseval} (see \emph{Appendix B: Evolution of Peak Angle}). Furthermore, the steady-state solution requires  $\frac{da_{\infty}(\theta)}{dt}=0$ so that \ref{WC} becomes
\begin{equation}
\label{WC2}
a_{\infty}(\theta)=g[h_{\infty}(\theta)].
\end{equation}
Thus, the shape of the activity profile at the steady state is equivalent to the net cortical input, cut off by $g$ at $\delta_{1}\equiv \theta_{c1}^{\star}$ and $\delta_{2}\equiv \theta_{c2}^{\star}$. Here, $\theta_{c1}^{\star}$ and $\theta_{c2}^{\star}$ are the critical cutoff angles for the steady-state activity profile, beyond which $a_{\infty}(\theta)$ would take on negative values. 

We emphasize that the values of the cortical parameters $J_{0}$, $J_{1}$, $c$, $T$, and $\beta$ are bounded by the physiological properties of V1. Varying these parameters in the subsequent analysis is therefore an investigation of their relative effects on hue processing, and we are not fine-tuning their weights to obtain specific hue tuning curves.

Here, we explore a range of values for the cortical and stimulus parameters under the constraint that the network activity remains between $0$ and $60$ spikes/sec, as motivated above. We further restrict $J_{1}>0$ and $J_{0}<0$ to elicit the local excitation and global inhibition connectivity ansatz of previous neural field models. Our main aim is to graphically characterize the relative effects of the parameters on the width,  $\Delta_{c}=\theta_{c2}^{\star}-\theta_{c1}^{\star}$, and peak height, $a_{\infty}(\bar{\theta})$, of the network tuning curves. Together, these two properties reflect the network selectivity and emergent signal strength, respectively. Note that these effects are robust to small additive white noise and may also be gleaned from the net input, expressed as in \ref{h_rewritten} and evaluated at the steady-state values of the coefficients.

It is also important to note here the difference between a network tuning curve and a single-neuron tuning curve. The former is a coarse-grained representation of the CO blob response, with the horizontal axis representing the gamut of hue preferences within a single network. A relatively large tuning width would therefore indicate considerable responses from a wide range of hue tuning cells and poor network selectivity. The single-neuron tuning curve, on the other hand, is an electrophysiological recording of an individual cell's response to a set of hue stimuli, with the horizontal axis representing the range of stimulus hue angles used in the experiment. The peak location of the single-neuron tuning curve would therefore indicate the hue preference of the individual neuron, while the width would represent its selectivity for that specific hue. Thus, though the two types of tuning curves are labeled and shaped similarly, the latter is only useful to characterize our network's constituent neurons and \emph{not} the emergent properties of the population as a whole \citep{Spherical}. 

\subsubsection{Roles of the Stimulus Strength and Cortical Threshold}
\begin{figure}[ht]
\centering
\includegraphics[width=.8\textwidth]{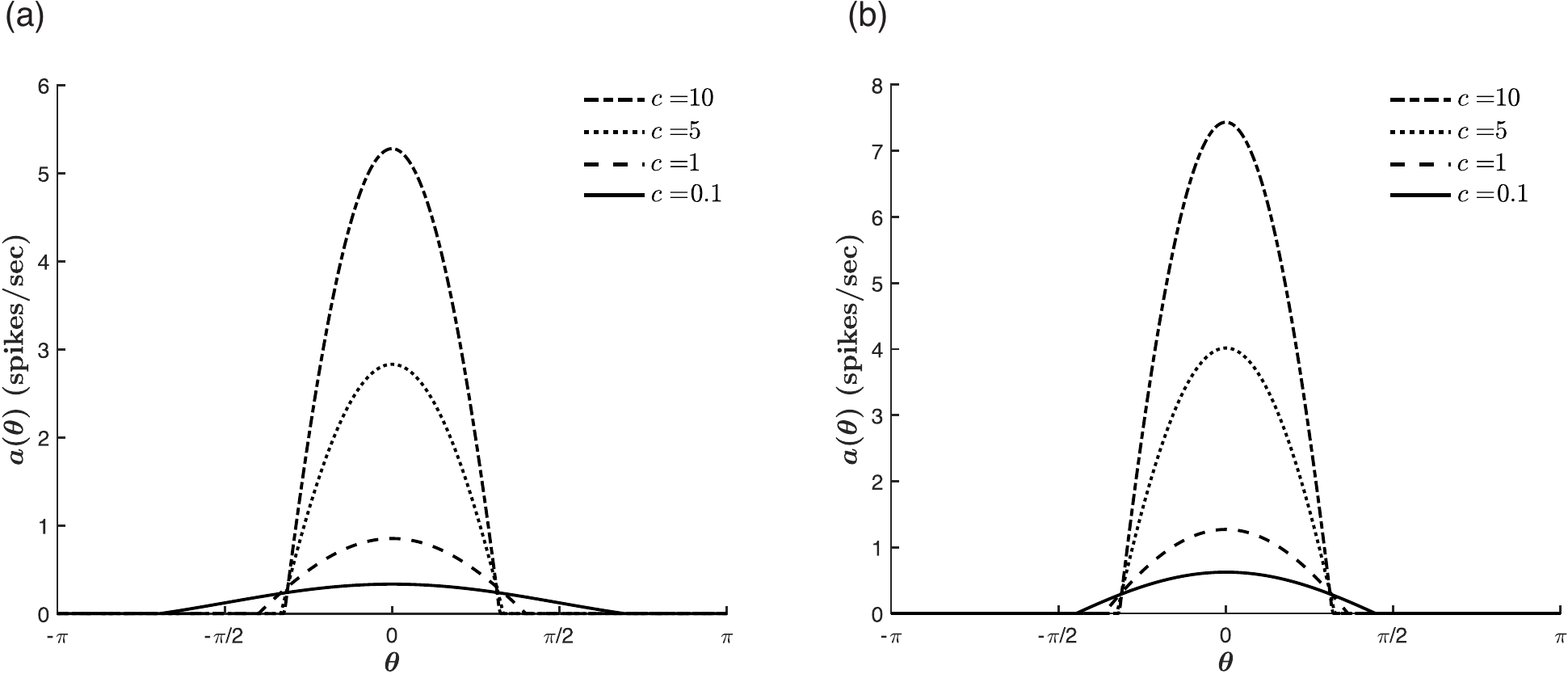}
\caption[Effect of $c$ on the Tuning Curve Properties]{Effect of $c$ on the tuning curve properties. The tuning role of $c$ quickly saturates, while its effect on the network response rate grows without bound. For $\bar{\theta}=0$, $\beta=1$, $T=-1$, and $J_{0}=-1$. (a) $J_{1}=0.2$ (b) $J_{1}=0.7$.}
\label{generalc}
\end{figure}

We begin by considering the role of the stimulus signal strength $c$ on the hue tuning width and peak height. Figure \ref{generalc} shows typical tuning curves for two values of $J_{1}$. We find that the stimulus strength has a quickly saturating effect on $\Delta_{c}$ for all $J_{1}>0$, which is more pronounced at lower values of $c$ as $J_{1} \rightarrow 0$. Above saturation, the  main contribution of the chromatic signal is to increase the network response, i.e., to increase $a_{\infty}(\bar{\theta})$. 

\begin{figure}[h]
\centering
\includegraphics[width=.8\textwidth]{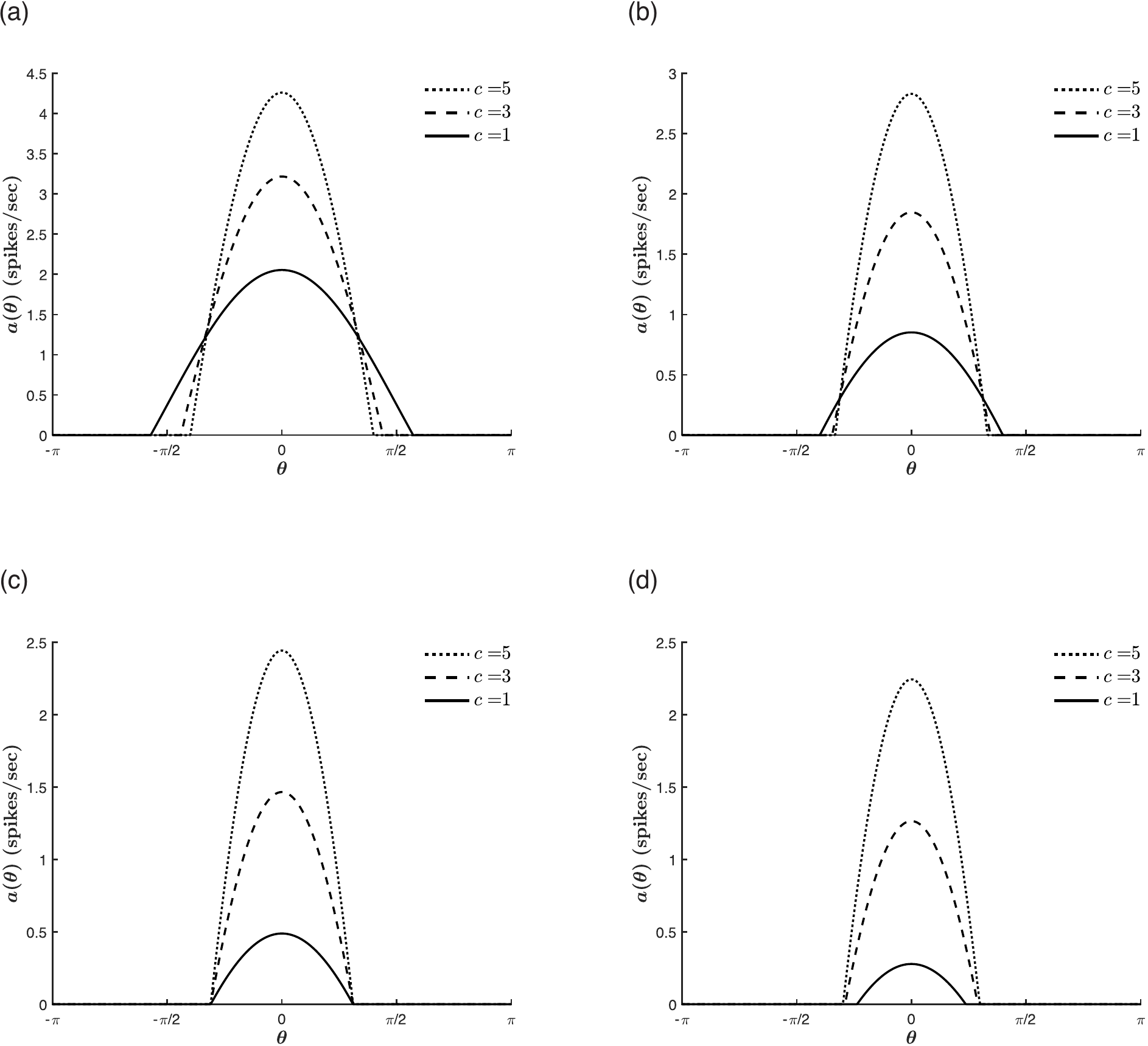}
\caption[Effect of $c$ for Varying Values of $T$]{Effect of $c$ on the tuning curve for varying values of $T$ with $\beta=1$, $J_{0}=-1$, $J_{1}=0.2$, and $\bar{\theta}=0$. Note that the small network response rates are due to the low values of $c$ chosen here. (a) $T=-5$. (b) $T=-1$. (c) $T=0$. (d) $T=0.5$.}
\label{Tiszero}
\end{figure}

We also note that at $T=0$, the trend reverses, such that increasing $c$ has no effect on the tuning width at $T=0$ and a widening effect for $T>0$. Figure \ref{Tiszero} illustrates this reversal with four tuning curves of matched parameters and varying values of $T$. The coupling of $c$ and $T$ must be considered because some neural field models (see \cite{Carandini, Amari, Dayan}) take $T=0$ for mathematical simplicity. Indeed, we might expect that there is no more physiological significance to choosing a threshold potential of $T=0$ mV than any other value, beyond their relative magnitudes to $h(\theta,t)$. However, the independence of $c$ and $\Delta_{c}$ at $T=0$ and the significance of the relative signs of $c$ and $T$ elsewhere suggest quite the opposite. The effect of the chromatic input on tuning the network hue selectivity weakens not only once the anisotropic strength parameter, $J_{1}$, is large enough to predominate, but also as $T \rightarrow 0$. 
\vskip 1cm

\onecolumngrid
\vskip 0.5cm

\begin{figure}[h]
\centering
\vspace{.1cm}
\includegraphics[width=.35\textwidth]{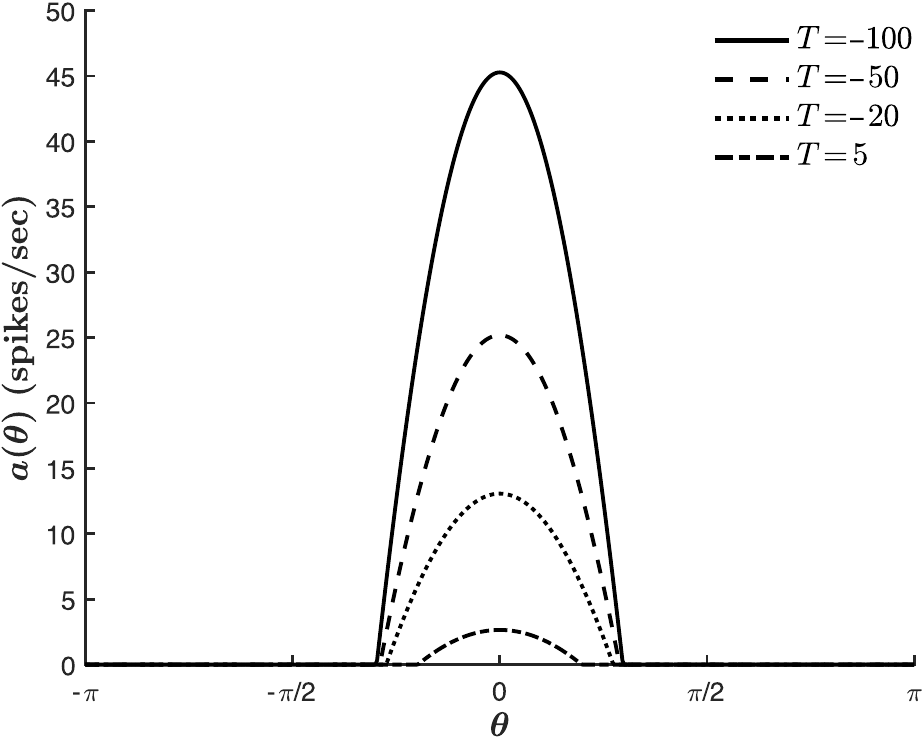}
\caption[Effect of $T$ on the Tuning Curve Properties]{Effect of $T$ on the tuning curve properties. $\bar{\theta}=0$, $\beta=1$, $J_{0}=-3$, $J_{1}=2$, and $c=10$. }
\label{Teffects}
\end{figure}

\twocolumngrid

The coupling of $c$ and $T$ is equally significant to the effects of $T$ on the tuning curve properties. Figure \ref{Teffects} shows that below a certain value, $T$ primarily modulates $a_{\infty}(\bar{\theta})$. However, for comparable magnitudes of the stimulus strength and threshold, $\lvert c \rvert \sim \lvert T \rvert$, we see a transition in which $T$ also begins to sharpen the tuning curve and continues to do so until the threshold surpasses $h(\theta,t)$ for all $\theta$ (i.e., for $\delta_{1}^{\star}=\delta_{2}^{\star}=0$). Accordingly, for higher stimulus strengths, the thresholding nonlinearity plays a greater role in modulating the network selectivity at lower and a wider range of $T$ values. 

\subsubsection{Roles of the Cortical Weights}
\begin{figure}[h]
\centering
\includegraphics[width=.85\textwidth]{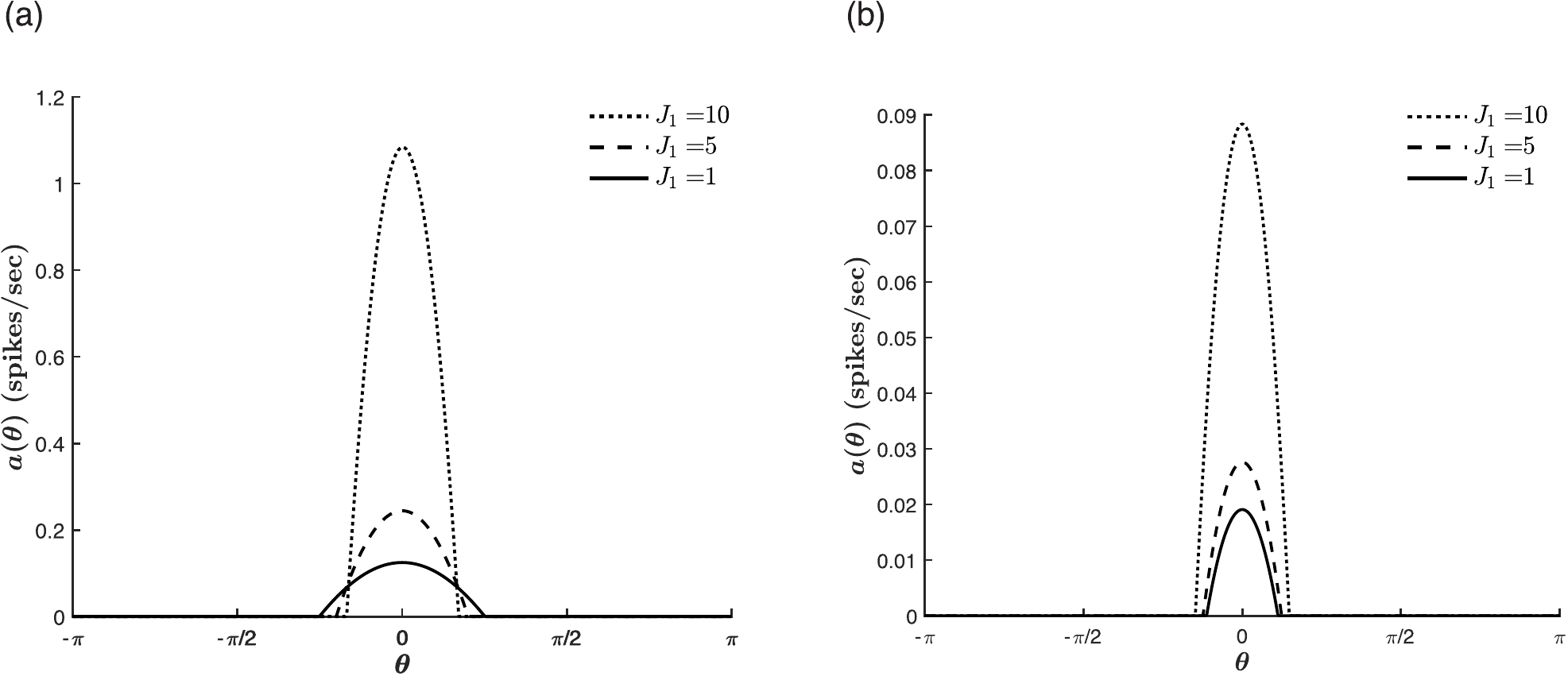}
\caption[Effect of $J_{1}$ for Varying Values of $T$]{Effect of $J_{1}$ on the tuning curve properties for varying values of $T$. $\beta=1$, $c=0.3$, $J_{0}=-10$, and $\bar{\theta}=0$. (a) $T=-1$. (b) $T=0.2$.}
\label{TandJ1}
\end{figure}

The anisotropic connectivity strength $J_{1}$ exhibits similar relationships to the tuning curve properties to those of $c$. That is, for $T<0$, $a_{\infty}(\bar{\theta})$ grows and $\Delta_{c}$ narrows with increasing $J_{1}$  (see Fig. \ref{TandJ1}a). The trend with respect to $\Delta_{c}$ reverses for $T>0$ (Fig. \ref{TandJ1}b), whereas the trend with respect to $a_{\infty}(\bar{\theta})$ remains unaffected.

\begin{figure}[h]
\centering
\includegraphics[width=.85\textwidth]{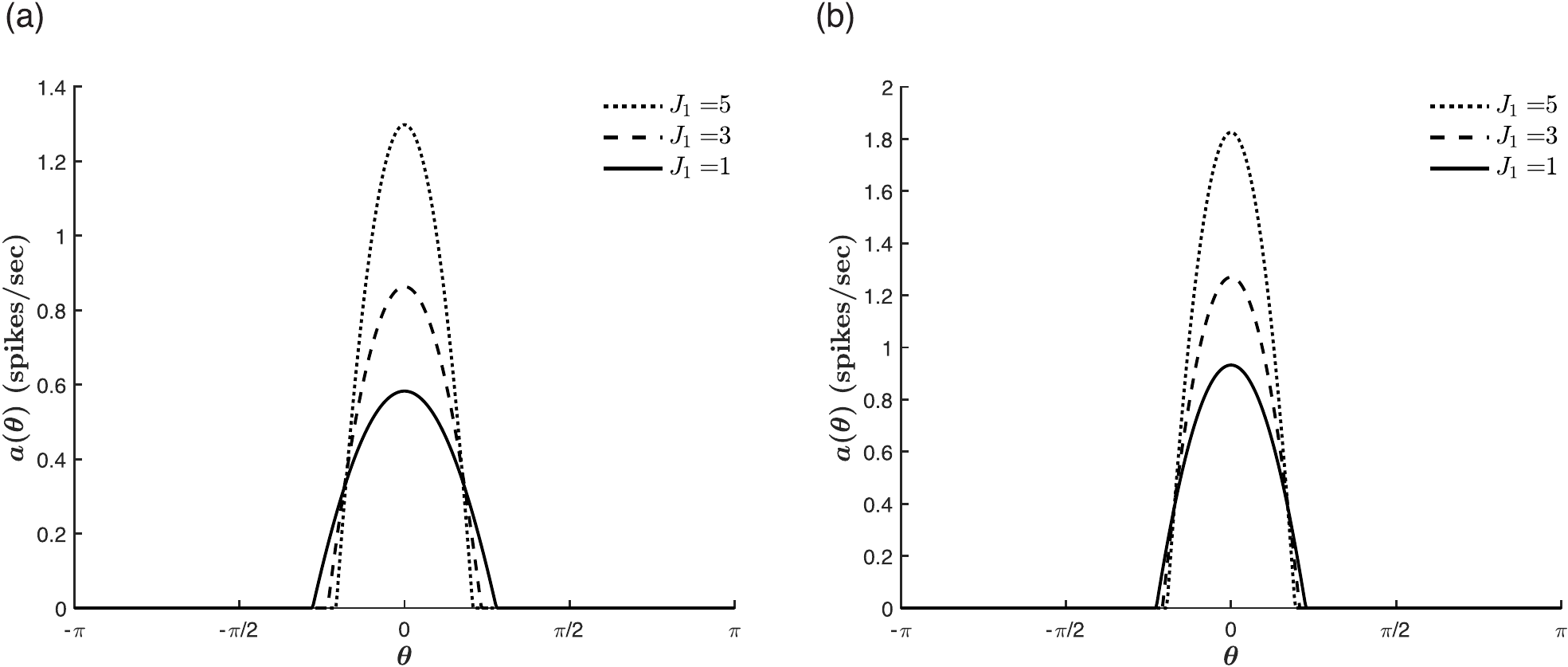}
\caption[Effect of $J_{1}$ for Varying Values of $c$]{Effect of $J_{1}$ on the tuning curve properties for different stimulus strengths. $\beta=1$, $T=-5$, $J_{0}=-9$, and $\bar{\theta}=0$.  (a) $c=1$. (b) $c=3$.}
\label{J1effects}
\end{figure}

These similarities are a mark of the competition between the external input and the cortical parameters in driving the network selectivity and reflect the fact that both parameters modulate the anisotropic terms of the model. This means that the role of $J_{1}$ in driving network selectivity becomes more significant with decreasing stimulus strength (see Fig. \ref{J1effects}). However, a large external input does not suppress the contribution of $J_{1}$ to the overall network activity. That is, increasing $J_{1}$ results in raising $a_{\infty}(\bar{\theta})$, regardless of the strength of the stimulus. Similarly, a relatively large value of $J_{1}$ does not restrict the growth of the network response with increasing stimulus strength. Thus, the anisotropic tuning introduced by the external input and the recurrent interactions act cooperatively to raise the network's response to the stimulus hue, and competitively to tune its selectivity.

\begin{figure}[h]
\centering
\includegraphics[width=.95\textwidth]{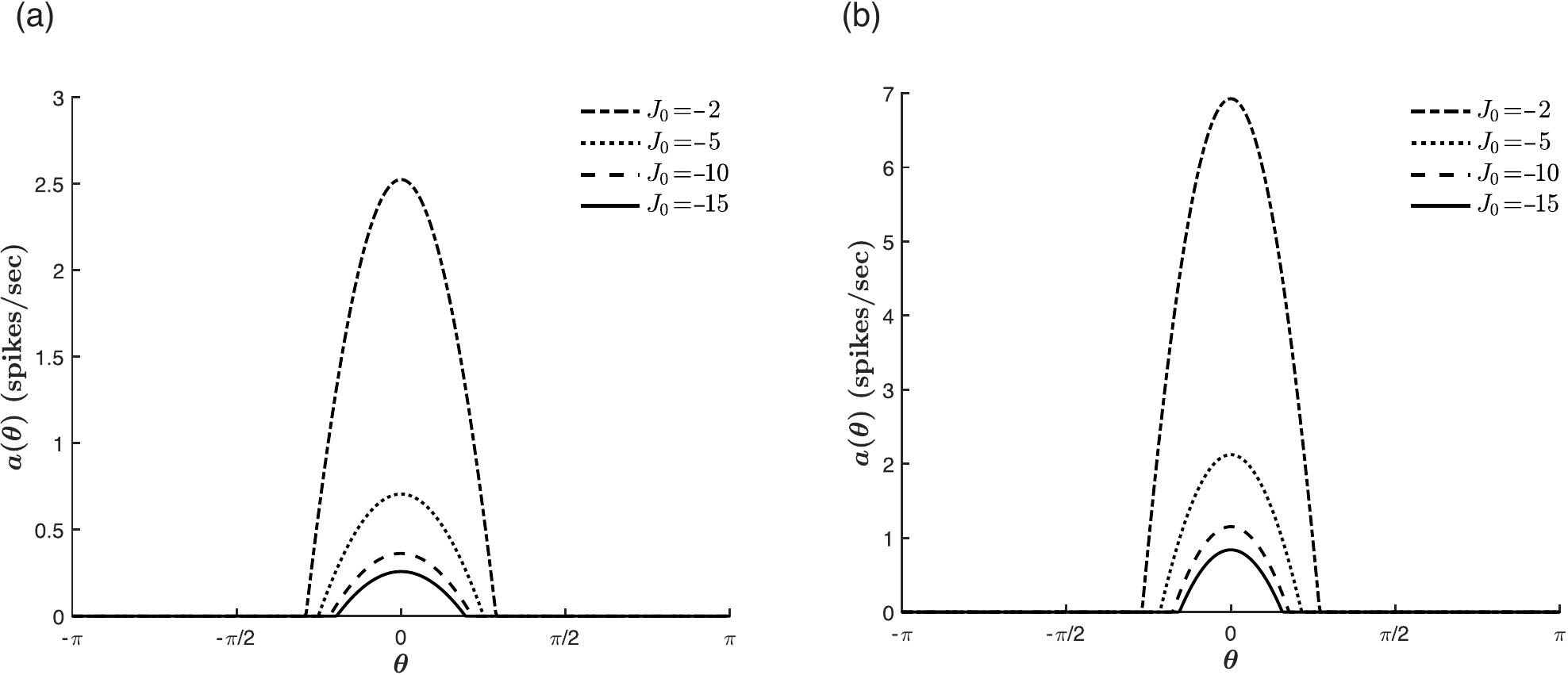}
\caption[Effect of $J_{0}$ for Varying Values of $c$]{Effect of $J_{0}$ on the tuning curve properties for varying stimulus strengths. $\beta=1$, $T=-2$, $J_{1}=2$, and $\bar{\theta}=0$. (a) $c=1$. (b) $c=6$.}
\label{J0effects}
\end{figure}

\begin{figure}[h]
\centering
\includegraphics[width=.95\textwidth]{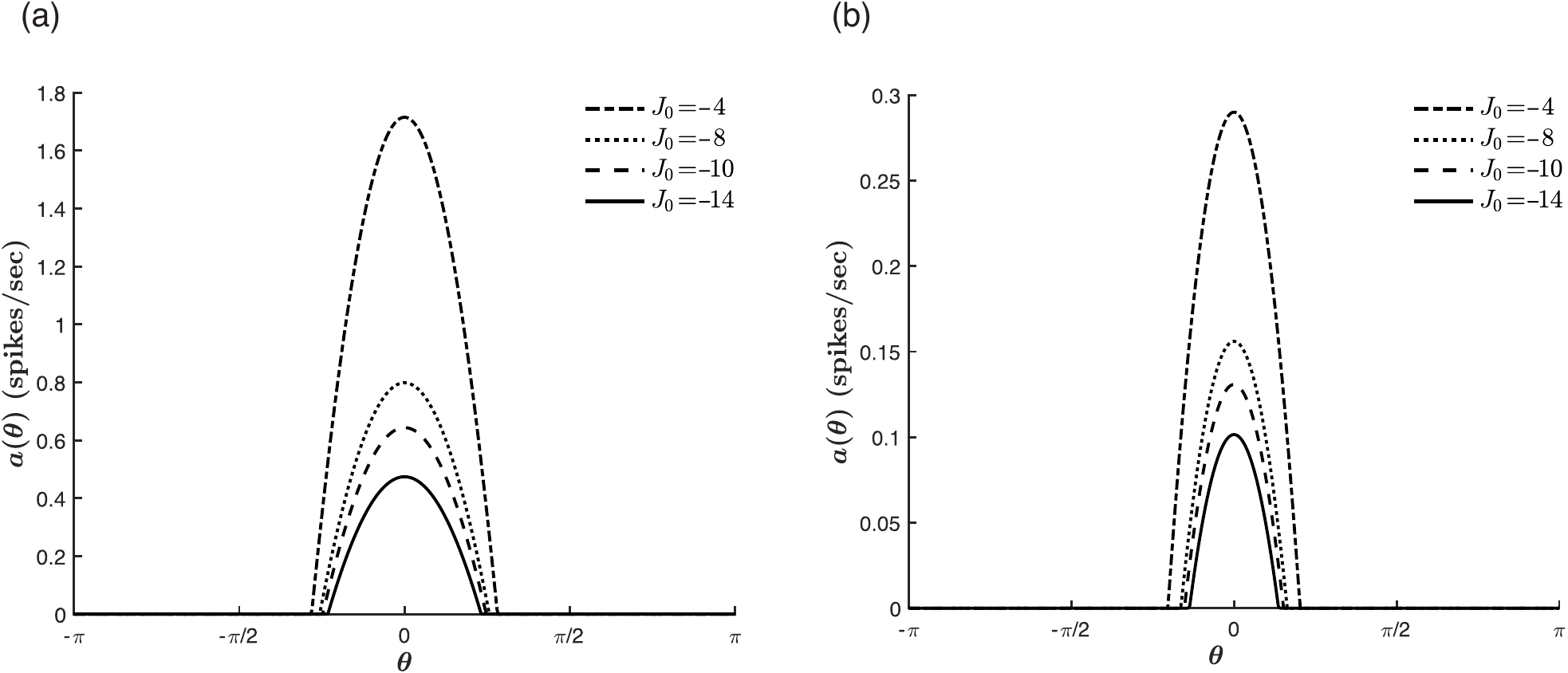}
\caption[Effect of $J_{0}$ for Varying Values of $T$]{Effect of $J_{0}$ on the tuning curve properties for varying values of $T$. $\beta=1$, $c=1$, $J_{1}=2$, and $\bar{\theta}=0$. (a) $T=-5$. (b) $T=0.2$.}
\label{J0andT}
\end{figure}

In contrast, $J_{0}$ acts cooperatively with the external stimulus to sharpen the curves. As shown in Fig. \ref{J0effects}, the tuning curves narrow with decreasing values of $J_{0}$, i.e., with an increase in the relative strength of global inhibition to global excitation, a trend which is conserved for various stimulus strengths. Furthermore, there is no trend reversal at $T=0$. Rather, for much of the parameter space, $J_{0}$ acts with the thresholding to sharpen the tuning curves, as is illustrated in Fig. \ref{J0andT}. This could be expected from the fact that at each point throughout the dynamics, both $T$ and $J_{0}$ act isotropically on all hue preferences, lowering or raising the input for all contributing neurons. However, this commonality also means that for $\lvert T \rvert >>\lvert c\rvert$ (where the effect of $T$ on $\Delta_{c}$ saturates, as explained above), the thresholding suppresses the role of $J_{0}$, analogous to the competition between $c$ and $J_{1}$. Finally, figures \ref{J0effects} and \ref{J0andT} also show that increasing the global inhibition acts to reduce the value of $a_{\infty}(\bar{\theta})$ for all $c$ and $T$.

We thus conclude that the emergent hue curves in V1 are \emph{both} inherited from the LGN \emph{and} built on the recurrent interactions. The competition between $J_{1}$ and $c$ points to a continuum of regimes in which either $h_\text{lgn}$ or $h_\text{ctx}$ dominates. However, in all regimes, $J_{0}$ works cooperatively with $c$ to narrow the curves, and all the parameters work together to raise the network response. Likewise, the competition between $J_{0}$ and $T$ (both cortical parameters) is modulated by the value of $c$, and the location of the peak is always completely determined by the LGN signal, regardless of the relative magnitudes of the cortical and stimulus strength parameters (see \emph{Appendix B: Evolution of Peak Angle}).

\onecolumngrid

\twocolumngrid

\subsubsection{Comparisons with the Orientation Tuning Ring Model}
\label{Comparison}
Finally, we seek to compare the emergent properties of the hue tuning model with those of the orientation tuning ring model \citep{Ben-Yishai, Hansel}. This leads us to separate the analysis into two regions: one corresponding to the \textbf{analytical regime} with $J_{0}<\frac{1}{2\pi \beta}$ and $J_{1}<\frac{1}{\pi \beta}$, and the other to the \textbf{extended regime} with $J_{1}\geq\frac{1}{\pi \beta}$ and $J_{0}$ constrained as described in the section \emph{Stability Analysis}. As detailed in \emph{Appendix A: Linear Solution}, the former regime defines the $(J_{0},J_{1})$ parameter space wherein the model permits a closed-from stable solution for cases in which the input into all cells is above threshold. By contrast, the dynamics in the latter, extended regime always implement thresholding and thus do not permit the linear closed-form solution. For comparison purposes, note that these parameter regimes are analogous to the orientation model's homogeneous and marginal regimes, respectively, labels which refer to the system's responses to unoriented stimuli.

An important difference between our two models is our choice to assume modularity for the color vision pathway. As described above, there is no consensus as to when and how the various visual features are separated along the visual pathway. That is, we do not yet understand how the brain recognizes the extent to which an activated color- and orientation-preferring neuron is signalling for a stimulus's color or orientation. And moreover, we do not know at which point of the visual pathway the differentiation becomes important. We have therefore chosen to emphasize the unoriented color selective cells localized in the CO blob regions of V1, though the model is intended to describe the color processing pathway broadly, for any color-preferring neurons regardless of other feature tuning capabilities. Thus, the choice of modularity is not to reject the possibility of joint feature processing, but rather to parse out the color mechanism for a separate analysis. Furthermore, it is in keeping with perceptual studies which indicate that the red-green and blue-yellow color-opponent systems are only responsive to color stimuli and not to broadband, white light \citep{Stockman}. The difference between our two models thus comes to our choice to consider the purely chromatic component of the input afferent from the LGN, whereas the orientation model incorporates external inputs with varying degrees of anisotropy, i.e.,
\begin{equation}
    h^{\text{ext}}(\theta)=c[1-\epsilon+\epsilon\cos(2\theta)] \ , \ 0 \leq \epsilon \leq 1/2
\end{equation}
where $\epsilon$ represents the degree of anisotropy. 

The differing assumptions underlying the formulation of $h(\theta,t)$ have important implications for the subsequent parameter analyses adopted by our two models. In the orientation tuning model, the authors detail the pronounced shift in the relative roles of the cortical and stimulus parameters in narrowing the tuning curve. In this setup, for $\epsilon \rightarrow 0.5$, an increase in $c$ widens the tuning curve, whereas for $\epsilon \rightarrow 0$, the tuning curve selectivity is completely determined by the cortical parameters. The latter scenario constrains the value of the analogous anisotropic cortical parameter, $J_{2}$, to the \emph{marginal} regime. 

In contrast, our model does not apportion separate regions of the parameter space to external and recurrent mechanisms. Rather, in both the analytical and extended regimes, the roles of $c$ and $J_{1}$ exist on a spectrum, where the effect of each parameter is suppressed by larger values of the other. Of course, this suppression is more stark in the extended regime because it covers larger values of $J_{1}$. In this sense, our color model draws a similar conclusion to that of the orientation model: when the anisotropic tuning provided by the recurrent interactions is large, the tuning from the stimulus is negligible, and vice versa. However, we emphasize that the transition is not sharp and that $c$ does have an effect on the tuning curve selectivity in the extended regime (see Fig. \ref{generalc}b), as does $J_{1}$ in the analytical regime.

In this regard, the two models are more consistent in their interpretations of $J_{0}$'s contribution to the selectivity of the tuning curves. That is, in the two regimes of each model, the inhibition acts cooperatively with the thresholding to sharpen the tuning curves. Here again, the orientation model makes a distinction between the marginal phase (i.e., $\epsilon=0$ and $J_{2}$ $\in$ marginal regime), wherein the tuning curve width is completely determined by the cortical anisotropy, and all other cases, where the isotropic inhibition and stimulus come into play. For these cases, the authors argue, $J_{0}$ does not act alone to narrow the curve: though $J_{0}$ may sharpen the tuning curves, it is the anisotropy from the input or cortical interactions which acts as the source of the orientation selectivity. Although our color model's tuning mechanism, too, requires a source of \emph{anisotropy}, we have emphasized above that there is no single source of \emph{hue selectivity}. When $J_{1}$ is small, in either regime, both the stimulus and the uniform inhibition are significant to the hue tuning mechanism. 

Ultimately, the orientation model sets up a dichotomy between two specific regions of parameter space. In the non-marginal case, $c$ is the primary player in the tuning mechanism, and in the marginal case, this role belongs to $J_{2}$. The uniform inhibition is thus given a secondary ``sharpening'' role. By contrast, in choosing a fully anisotropic $h_{\text{lgn}}$, our color model does not encompass an analogous marginal phase with an always-dominating $J_{1}$. Rather, even for large $J_{1}$, the uniform inhibition is at least equally important to the modulation of the tuning width. In fact, as we have shown above, for larger values of $c$, $J_{0}$ is more effective than $J_{1}$ in narrowing the tuning curves, for both the analytical and extended regimes. 

We thus stress that the two  regimes, though analogous to those of the orientation model, do not constitute a division in the hue processing mechanism. Rather, we define the boundary between the analytical and extended regimes solely by whether or not the linear case exists. It is therefore determined by the values of $J_{0}$ and $J_{1}$ for which the linear solution applies, given that the values of $c$, $T$, and $\beta$ keep the input above threshold throughout the dynamics (\emph{Appendix A: Linear Solution}). We note that for each combination of $J_{0}$ and $J_{1}$ within the analytical regime there exists also a nonlinear case, in which $h(\theta,t)$ is cut off by the thresholding nonlinearity and, thereby, the linear solution does not apply. Our definition differs from that of the orientation model, which demarcates the boundary between the homogeneous and marginal phases based on the emergent steady-state tuning curves alone. For more on this approach, see the discussion of the broad and narrow profiles in \cite{Hansel}. As we will show next, the boundary is integral to the corresponding stability analysis of the steady-state tuning curves.

\subsection{Stability Analysis}
\label{stabilityanalysis}
To analyze the stability of the emergent tuning curves, we turn once more to our separable activity ansatz assumed in the eigenfunction decomposition of equation \ref{decompose}. This means that we are faced again with a nonlinearity-induced coupling of the time-dependent coefficients and, consequently, the analytical intractability of the associated stability analysis. We therefore set up the Jacobian matrix for a numerical analysis of the local stability. 

We begin by adding a small perturbation of the form
\begin{equation}
\label{pertubationexpansion}
    \delta a(\theta,t)=\sum \limits_{\mu} D_{\mu}(t){\hat{e}_{\mu}(\theta)}
\end{equation}
and substituting the resulting activity into \ref{WC}. The eigenmodes then evolve according to the following equation for the coefficients $D_{\mu}$ (see \emph{Appendix C: Linear Stability Analysis}): 
 \begin{align}
 \begin{split}
 \label{nonlinearstability}
 \tau_{0}\frac{dD_{\nu}(t)}{dt}=-D_{\nu}(t) &+\beta\int_{\delta_{1}^{\star}}^{\delta_{2}^{\star}}\big[\delta q_{0}(t)+\delta q_{\text{{R}}}(t)\cos(\phi) \\
    &+\delta q_{\text{{I}}}(t)\sin(\phi)\big]\hat{e}^{*}_{\nu}(\phi)d\phi, 
 \end{split}
 \end{align} 
where $\delta_{1}^{\star}$ and $\delta_{2}^{\star}$ are the critical cutoff angles of the steady-state solution, obtained numerically. We observe that the integrand of \ref{nonlinearstability} is a function of $D_{0}$, $D_{\text{-}1}^{R}$, and $D_{\text{-}1}^{I}$ alone, and, as such, the stability of the steady-state tuning curve is completely determined by the stability of these first-order coefficients.

Evaluating the integrals for $\nu=0$ and $\nu=-1$, and noting from \ref{subs} that
\begin{gather}
\delta q_{0}=\sqrt{2\pi}J_{0} \delta c_{0} \equiv \sqrt{2\pi}J_{0} D_{0} \nonumber \\
\delta q_\text{{R}}=\sqrt{2\pi}J_{1} \delta c_{\text{-}1}^{R} \equiv \sqrt{2\pi}J_{1} D_{\text{-}1}^{R} \nonumber \\  
\delta q_\text{{I}}=\sqrt{2\pi}J_{1} \delta c_{\text{-}1}^{I} \equiv \sqrt{2\pi}J_{1} D_{\text{-}1}^{I},
\end{gather}
we obtain the following system of equations for the evolution of the characteristic coefficients: 
\begin{align}
    \begin{split}
    &\tau_{0}\frac{dD_{0}}{dt}= \big[\beta J_{0}(\delta_{2}^{\star}-\delta_{1}^{\star})-1 \big]D_{0}\\
    &\qquad+\big[\beta J_{1}(\sin\delta_{2}^{\star}-\sin\delta_{1}^{\star}) \big]D_{\text{-}1}^{R}\\
    &\qquad +\big[\beta J_{1}(\cos\delta_{1}^{\star}-\cos\delta_{2}^{\star}) \big]D_{\text{-}1}^{I}
     \end{split} \nonumber \\[1ex]
    \begin{split}
     &\tau_{0}\frac{dD_{\text{-}1}^{R}}{dt}=\big[\beta J_{0}(\sin\delta_{2}^{\star}-\sin\delta_{1}^{\star})\big]D_{0}\\ 
     &\qquad +\big[\tfrac{\beta J_{1}}{4} \big( 2\delta_{2}^{\star}-2\delta_{1}^{\star}
     +\sin(2\delta_{2}^{\star})-\sin(2\delta_{1}^{\star})\big)
      -1 \big]D_{\text{-}1}^{R}\\
      &\qquad + \big[ \tfrac{\beta J_{1}}{4} \big(\cos(2\delta_{1}^{\star})-\cos(2 \delta_{2}^{\star})\big)\big]D_{\text{-}1}^{I}
    \end{split} \nonumber \\[1ex]
     \begin{split}
     &\tau_{0}\frac{dD_{\text{-}1}^{I}}{dt}=\big[\beta 
      J_{0}(\cos\delta_{1}^{\star}-\cos\delta_{2}^{\star}) \big]D_{0}\\
    &\qquad +\big[\tfrac{\beta J_{1}}{4} \big(\cos(2\delta_{1}^{\star})-\cos(2 \delta_{2}^{\star})\big)\big]D_{\text{-}1}^{R}\\
     &\qquad+\big[\tfrac{\beta J_{1}}{4} \big(2 \delta_{2}^{\star}-2\delta_{1}^{\star}+\sin(2\delta_{1}^{\star})-\sin(2\delta_{2}^{\star})\big)-1 \big]D_{\text{-}1}^{I} 
    \end{split} 
\end{align}

The entries of the corresponding Jacobian matrix consist of the bracketed prefactors, and may equally be obtained from the general system of equations for the global network dynamics, as follows: 
\begin{equation}
 \mathlarger{\mathbb{J}}=
\begin{bmatrix}
  \dfrac{\partial f_1}{\partial c_{0}} & 
    \dfrac{\partial f_1}{\partial c_{\text{-}1}^{R}} & 
    \dfrac{\partial f_1}{\partial c_{\text{-}1}^{I}} \\[3ex] 
  \dfrac{\partial f_2}{\partial c_{0}} & 
    \dfrac{\partial f_2}{\partial c_{\text{-}1}^{R}} & 
    \dfrac{\partial f_2}{\partial c_{\text{-}1}^{I}} \\[3ex]
  \dfrac{\partial f_3}{\partial c_{0}} & 
    \dfrac{\partial f_3}{\partial c_{\text{-}1}^{R}} & 
    \dfrac{\partial f_3}{\partial c_{\text{-}1}^{I}}
\end{bmatrix}_{\mathlarger{c}_{0}^{\star}, \; \mathlarger{c}_{\text{-}1}^{R^{ \star}}, \; \mathlarger{c}_{\text{-}1}^{I^{\star}}}
\end{equation}
where $\mathlarger{f}_{1}$, $\mathlarger{f}_{2}$, and $\mathlarger{f}_{3}$ are the right-hand sides of the equations in \ref{systemofequationseval} and the first-order partial derivatives are evaluated at the steady-state values of $c_{0}$, $c_{\text{-}1}^{R}$, and $c_{\text{-}1}^{I}$. The stability of the tuning curve is then determined by the eigenvalues of $\mathbb{J}$. 

We note that the existence of a steady state is a function of the cortical strengths $J_{0}$ and $J_{1}$. By fixing the values of $\beta$, $c$, $\bar{\theta}$, and $T$, we are left with a two-parameter family of differential equations, allowing us to analyze this dependence numerically in the associated $(J_{0},J_{1})$ parameter space. 

Carrying out a parameter sweep across this space, we find that the system features a bifurcation curve, below which the model permits steady-state solutions and above which no equilibrium exists. To determine stability within the former region, we compute $\mathbb{J}$ at the emergent steady-state tuning curves of various points in the parameter space. Solving the associated characteristic equations, we observe that the eigenvalues are always real and negative, and thus conclude that \textbf{all emergent steady-state tuning curves are stable}.

\begin{figure}[ht]
\centering
\includegraphics[width=.45\textwidth]{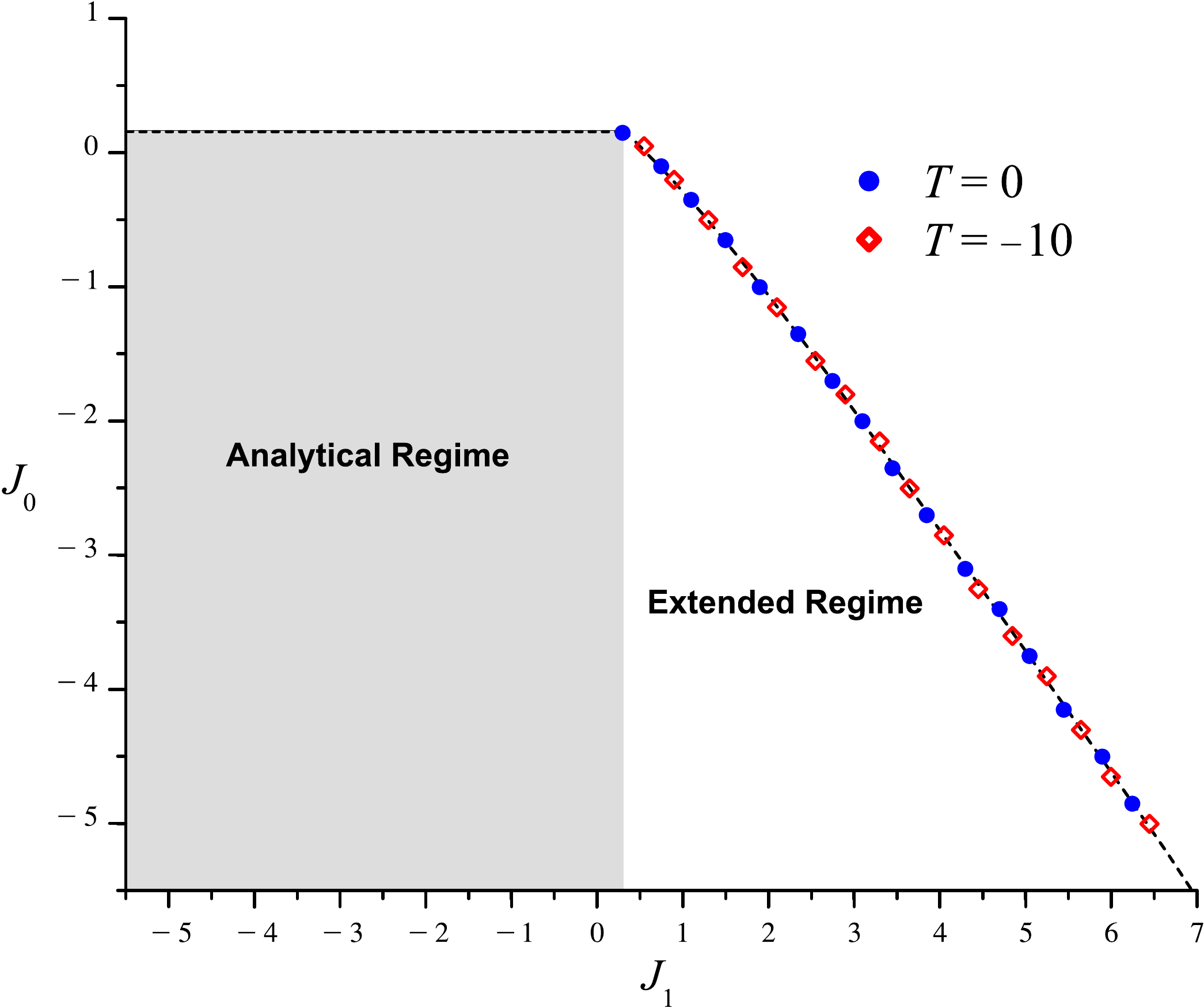}
\caption[Bifurcation Diagram in the $(J_{0},J_{1})$ Parameter Space]{Bifurcation diagram for $\beta=1$ and $c=1$ for two values of threshold (shown in the legend). The grey and white regions correspond to the analytical and extended regimes, respectively. The black dashed line is the bifurcation curve, above which the tuning curves grow without bound. The overlaid symbols correspond to points tested in a parameter sweep over the extended regime. Notably, the parameter sweep produces the same bifurcation curve for both values of $T$. Here, we must note that for critical values of $T$, for which the input is not large enough to generate activity, the model permits the trivial $a(\theta)=0$ steady-state solution in both the analytical and extended regimes. This solution, however, is unstable to perturbations large enough to make the input cross the threshold. For more on bifurcation theory in the context of neural fields, see \cite{Gross, Hesse}. See also \cite{Hansel} for an analogous ``phase diagram'' analysis of the orientation ring model.}
\label{stability}
\end{figure}

Figure \ref{stability} shows the bifurcation diagrams for two families of equations, distinguished by their values for $T$. Most notable is the \emph{extended regime}, which permits stable steady-state solutions beyond the boundary set by the linear case (\emph{Appendix A: Linear Solution}). As this parameter regime is not accessible to the linear solution, the tuning curves in this regime are necessarily a product of the thresholding nonlinearity and are thus always cut off below $\lvert \theta \rvert =\pi$. The thresholding nonlinearity therefore not only expands the region of stability, but also ensures that the tuning curves emerging within the extended regime are selective for hue. As we have seen, this expansion is pivotal when the external input is weak and the anisotropic cortical strength plays the larger role in narrowing the tuning curves. Furthermore, regardless of input strength, it allows for a larger overall network response, as the peak activity, $a_{\infty}(\bar{\theta})$, grows with increasing $J_{1}$. Finally, as we will see in the following section, in the absence of any stimulus (i.e., for $c=0$), the extended regime features the spontaneous generation of stable tuning curves and may thus serve as the bedrock for color hallucinations.

However, looking back at Fig. \ref{stability}, perhaps most striking is the horizontal portion of the bifurcation curve at $J_{0}=\frac{1}{2 \pi \beta}$ for $J_{1}<\frac{1}{\pi \beta}$, which sets the same stability conditions on $J_{0}$ and $J_{1}$ as in the linear case. This is despite the fact that many of the points in the analytical regimes of the two featured families correspond to solutions that implement thresholding, thus signifying that the analytical regime is not an exclusively \emph{linear} one. 

The key to understanding the shape of this region lies in noticing that the bifurcation diagram does not change for varying values of $c$, $T$, and $\bar{\theta}$, as shown in Fig. \ref{stability} for the two values of $T$. The stability conditions on $J_{0}$ and $J_{1}$ are thus uniquely determined by $\beta$ alone. Furthermore, for the general diagram (i.e., with $\beta$ fixed and $c$, $T$ unfixed), each point of the analytical regime permits linear solutions, in addition to the ones that implement thresholding. Accordingly, the uniqueness of the bifurcation diagram implies that at each point of the analytical $(J_{0},J_{1})$ subspace, the stability of the latter, nonlinear solutions is equivalent to that of the linear solutions. This means that the boundary at $J_{0}=\frac{1}{2 \pi \beta}$ set by the linear case (\emph{Appendix A: Linear Solution}) applies to the full, nonlinear model as well.

\subsection{A Turing Mechanism for Color Hallucinations}
\label{Turingsec}
\subsubsection{Biological Turing Patterns}
Underpinning our hue tuning model is the mathematics of reaction-diffusion systems, for which, in particular, Alan Turing's treatment of biological pattern formation offers many valuable insights \citep{Turing}. The general Turing mechanism assumes a system of two interacting chemicals, whose local reaction and long-range diffusion properties govern the dynamics of their relative concentrations. In the original framework these chemicals are termed ``morphogens'' to elicit their form-producing capabilities within a developing embryo, whose anatomical structure emerges as a result of their underlying concentration dynamics. This, for instance, may be attributed to the morphogens' catalysis of organ formation in different parts of the developing organism.

Most analogous to our model is the formulation which distributes the morphogens across a continuous ring of tissue, parameterized by the cellular position $\theta$. Assuming that the system never deviates far from the underlying homogeneous steady state, the two dynamical state equations for their concentrations, $X$ and $Y$, take the linear form
\begin{align}
    \frac{dX(\theta,t)}{dt}=aX(\theta,t)+bY(\theta,t)+D_{\scriptscriptstyle{X}}\nabla^{2}X(\theta,t) \nonumber \\
    \frac{dY(\theta,t)}{dt}=cX(\theta,t)+dY(\theta,t)+D_{\scriptscriptstyle{Y}}\nabla^{2}Y(\theta,t),
\label{Turingeq}
\end{align}
where $a$, $b$, $c$, and $d$ represent the chemical reaction rates, and $D_{\scriptscriptstyle{X}}$ and $D_{\scriptscriptstyle{Y}}$ are the diffusion rates of $X$ and $Y$, respectively. Here, we set $a$, $c>0$, so that increasing the concentration of $X$ activates the production of both $X$ and $Y$, and $b$, $d<0$ so that $Y$ has an inhibitory effect on the production of both chemicals \citep{Hoyle}. 

In the absence of diffusion (i.e., with $D_{\scriptscriptstyle{X}}=D_{\scriptscriptstyle{Y}}=0$), the system has a homogeneous steady-state solution, $(X,Y)=0$, whose stability is determined by a Jacobian composed of the reaction rates, $\left[\begin{smallmatrix}
a & b \\
c & d
\end{smallmatrix}\right]$, and hence by the system's local chemical properties alone \citep{Hoyle}. Note that at this point the system is circularly symmetric with respect to interchanging any two cells on the ring.

Assuming the existence of a stable steady-state solution, and the corresponding requirements on the rate parameters $a$-$d$, we next set the diffusive terms $D_{\scriptscriptstyle{X}}, D_{\scriptscriptstyle{Y}}>0$, taking the separable ansatz for the general solution:
\begin{align}
\label{Turingsolutions}
    X&=\sum \limits_{\mu=-\infty}^{\infty}A_{\mu}e^{\lambda_{\mu} t}e^{i \mu \theta} \nonumber \\[1ex]
    Y&=\sum \limits_{\mu=-\infty}^{\infty}B_{\mu}e^{\lambda_{\mu} t}e^{i \mu \theta}.
\end{align}
Furthermore, we set $D_{\scriptscriptstyle{X}}<D_{\scriptscriptstyle{Y}}$ to generate the local excitation and lateral inhibition of the morphogen concentrations \citep{Hoyle, Murray}, evoking the connectivity function ansatz \ref{differenceofcosines}. The underlying steady state then remains stable if the real parts of the eigenvalues $\lambda_{\mu}$, obtained from the modified Jacobian, are negative. With the reaction rates fixed from the stability conditions above, these eigenvalues are functions of the diffusion parameters alone. Thus, the conditions for stability may be thought of in terms of a bifurcation diagram in the $(D_{\scriptscriptstyle{X}}$, $D_{\scriptscriptstyle{Y}})$ phase space, comparable to Fig. \ref{stability}. 

From here, a set of additional conditions may be placed on $D_{\scriptscriptstyle{X}}$ and $D_{\scriptscriptstyle{Y}}$ so that the system undergoes a Turing bifurcation, wherein at least one $\lambda_{\mu}$ becomes positive and the homogeneous steady state loses its stability. With the addition of a small random perturbation, the instability results in the growth of the corresponding eigenmodes $e^{i \mu \theta}$, such that, over time, \ref{Turingsolutions} is dominated by the eigenmodes with largest $\lambda_{\mu}$. These represent stationary waves whose wavelengths are set by the circumference of the ring (i.e., by the spatial properties of the medium) and whose growth is bounded by the higher-order terms which had been initially ignored in the near-equilibrium formulation \citep{Scholz, Turing, Murray}. The underlying circular symmetry is thus broken and a spatial pattern is formed. 

In his seminal paper, Turing extrapolated this mechanism to explain various biological phenomena, such as the development of petals on a flower, spotted color patterns, and the growth of an embryo along various directions from an original spherical state. A hallmark of each of these examples is that there is no input into the system, so the emergent patterns reflect a mechanism of spontaneous symmetry breaking, onset by a perturbation of ``some influences unspecified'' \citep{Turing}. In light of this, we ask, can the visual cortex self-generate the perception of hue? 

\subsubsection{Spontaneous Symmetry Breaking \\and Color Hallucinations}

To assess our model's ability to self-organize in the absence of visual input, we set $c=0$ and seek to establish the presence of a Turing mechanism marked by the following three features: 
\begin{enumerate}
    \item A system comprised of local excitation and long-range inhibition.
    \item Spontaneous symmetry breaking in the absence of input within a region of a parameter space defined by the relevant bifurcation parameter(s).
    \item The emergence of patterns that are bounded by the system's nonlinearities.
\end{enumerate}

\begin{figure}[h]
\centering
\includegraphics[width=.4\textwidth]{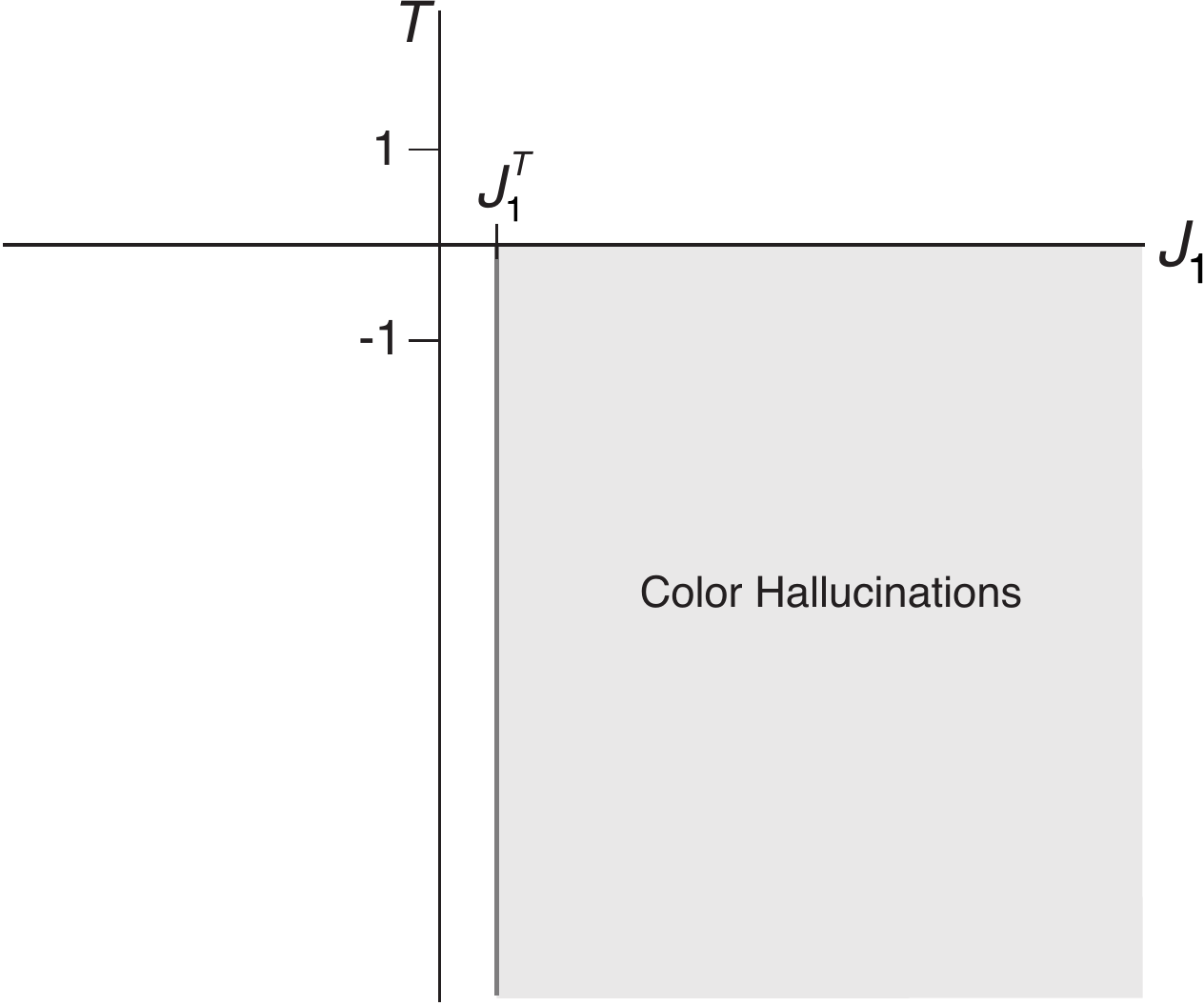}
\caption[The Onset of Color Hallucinations]{The onset of color hallucinations in the $(J_{1}, T)$ parameter space. The model generates spontaneous hue tuning curves beyond $J_{1}=J_{1}^{\mathsmaller{T}}\equiv \frac{1}{\pi \beta}$ and below $T=0$.}
\label{bifucationTuring}
\end{figure}

\begin{figure}[h]
\centering
\includegraphics[width=.48\textwidth]{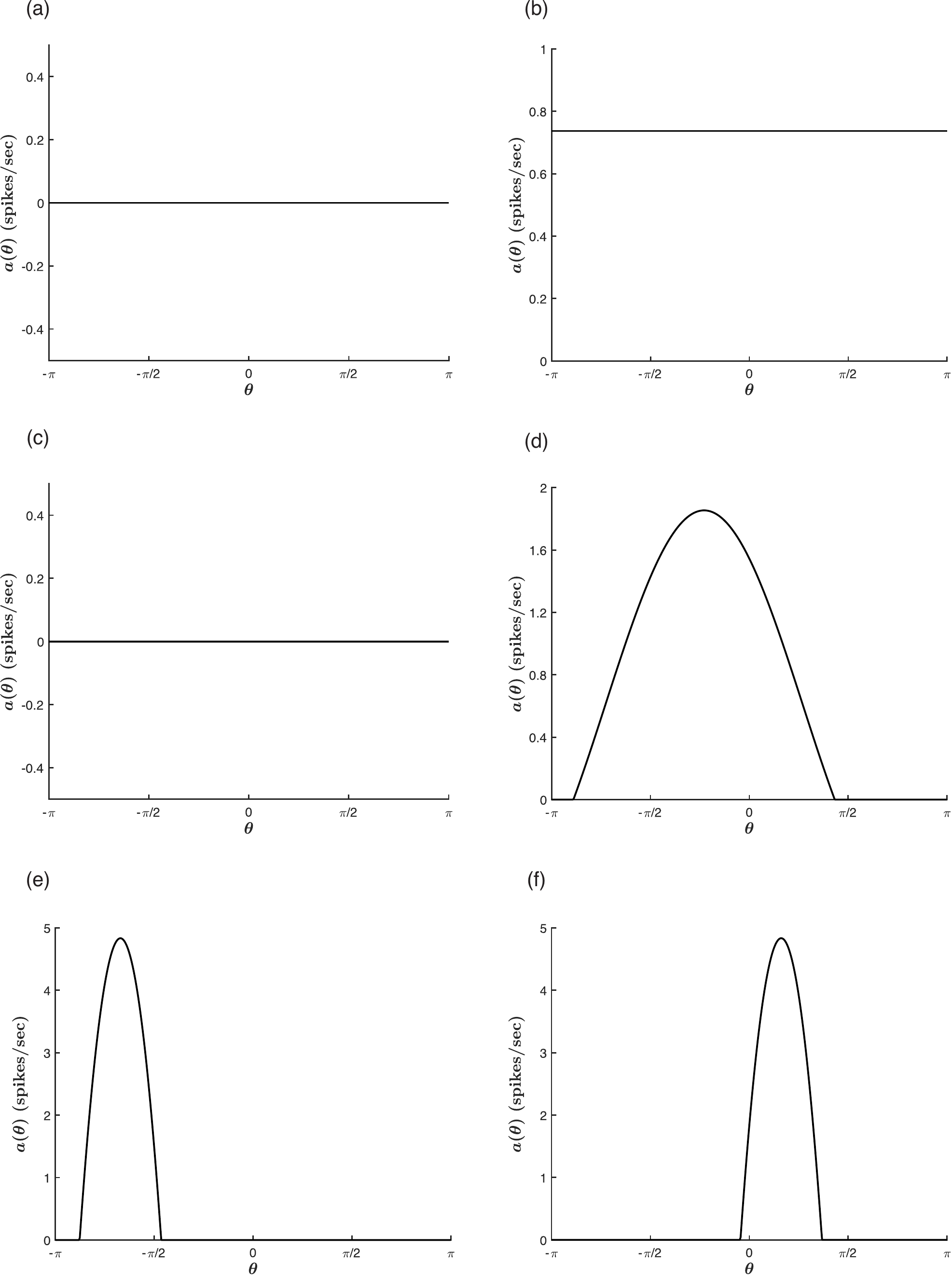}
\caption[Spontaneous Pattern Formation]{Spontaneous pattern formation in the absence of input ($c=0$). $\beta=1$. (a)-(b): $J_{0}=-2$, $J_{1}=0.1$ (a) $T=0$ (b) For $T<0$, the homogeneous steady-state value increases. Here, $T=-10$. (c)-(d): Pattern formation in the extended regime for $J_{0}=-2$, $J_{1}=0.4$. (c) No hue tuning curve emerges for $T\geq0$. Here, $T=0$. (d) $T=-10$. A hue tuning curve is generated in the absence of external input. (e)-(f): $T=-10$, $J_0=-7$, $J_{1}=6$. The emergent tuning curve is more selective for larger values of $J_{1}$. For each run, the activity is peaked about a different angle, set by the random initial conditions. The peak value and tuning width are consistent between trials.}
\label{Turing}
\end{figure}

As noted above, requiring $D_{\scriptscriptstyle{X}}<D_{\scriptscriptstyle{Y}}$ in \ref{Turingeq} sets up the diffusion-driven activator-inhibitor dynamics governing the evolution of the morphogen concentration across the ring of cells. With these assumptions, Turing's reaction-diffusion equations bear a strong resemblance to our one-population generalization of the excitatory and inhibitory color cell dynamics in the absence of LGN input: 
\begin{align}
\begin{split}
\label{noinput}
&\tau_{0}\frac{da(\theta, t)}{dt} =-a(\theta, t) \\
&+ g\left[\int_{-\pi}^\pi \left(J_{0}+J_{1}\cos( \theta-\theta')\right)a(\theta',t)d\theta' \right],
\end{split}
\end{align}
where the local excitation and long-range inhibition are incorporated in the anisotropic interaction term $J_{1}\cos( \theta-\theta')$, and the reaction terms $aX(\theta,t)$, $bY(\theta,t)$, $cX(\theta,t)$, and $dY(\theta,t)$ find their neural analogue in the term $-a(\theta,t)$. Importantly, the notions of ``local'' and  ``long-range'' here describe interactions in the DKL space, and not in the \emph{physical} cortical space correlate to Turing's ring of tissue. Accordingly, we treat $J_{1}$ as the Turing bifurcation parameter and look for spontaneous color tuning beyond a bifurcation point $J_{1}=J_{1}^{\mathsmaller{T}}$. Additionally, we observe that the onset of pattern formation is determined by a critical value of $T$, so that the relevant parameter space for our exploration is $(J_{1}, T)$ (Fig. \ref{bifucationTuring}). This analysis is summarized in Fig. \ref{Turing}.  

We observe that within the analytical regime, the system generates a stable homogeneous steady-state solution $a_{\infty}(\theta)=\text{const}\geq 0$ for all values of the parameters $\beta$, $T$, $J_{0}$, and $J_{1}$ (Fig. \ref{Turing}a-b). As such, from the closed-form linear steady-state solution (\emph{Appendix A: Linear Solution}), we obtain
\begin{equation}
          a_{\infty}(\theta)  = 
  \begin{cases}
  \mathlarger{-\frac{\beta T}{1-2\pi \beta J_{0}}}  & \text{  for } T\leq 0 \\
        \qquad \;\; \mathlarger{0} & \text{   for } T>0.
  \end{cases}
  \end{equation}

We further observe that beyond $J_{1}=\frac{1}{\pi\beta}$, a stable homogeneous steady-state solution remains at $a_{\infty}(\theta)=0$ for $T \geq 0$ (Fig. \ref{Turing}c). However, at $T=0$, this radial symmetry is broken, and the cortex generates spontaneous tuning curves with peak locations determined by the random initial conditions (Fig. \ref{Turing}d-f). Thus, the system bifurcates when $J_{1}=\frac{1}{\pi\beta}$ and $T=0$, permitting the onset of color hallucinations in a region defined by these two values (Fig. bifucationTuring). Note that the unimodal tuning curves predict stationary, single-hued phosphenes. Extension to other CO blob networks would therefore indicate a hallucination comprised of multiple phosphenes of varied hues, each determined by the local cortical activity at hallucinatory onset.

Bearing these predictions in mind, we point to a recent functional MRI study of blind patients experiencing visual hallucinations \citep{Hahamy}. The study attributes these visions to the activation of the neural networks underlying normal vision, precipitated by the hyper-excitability of the cortex to spontaneous resting-state activity fluctuations when it is deprived of external input. This is suggestive of the required lowering of neuronal threshold at the onset of color hallucinations predicted here. Notably, a reduction in membrane potential threshold has also been attributed to the action of hallucinogens \citep{hallucthreshold1, hallucthreshold2}. 

Finally, we note that the stability of the emergent tuning curves is determined by the bifurcation diagram of Fig. \ref{stability}. This means that, in addition to expanding the region of stability in the presence of chromatic stimuli, the model's nonlinearity allows for \emph{stable}, spontaneous color hallucinations in their absence.

Having thus established a Turing-like mechanism for our model's self-organization, we end with an analogy to Turing's original diffusion-driven formulation. In his concluding example, Turing applies the mechanism to explain the growth of an embryo along various axes of its original spherical state. This growth is driven by diffusion, directed by the ``disturbing influences,'' shaped by the system's chemical and physical properties, and bounded by the system's nonlinearities. It is all too clear to see the parallels with our hue tuning model, wherein a hallucination is driven by the anisotropy of the cortical interactions, its hue determined by the initial conditions, its selectivity shaped by the cortical parameters, and its stability ensured by the thresholding nonlinearity. 

\section{Discussion}\label{sec12}
This paper presents a neural field model of color vision processing which reconceptualizes the link between chromatic stimuli and our perception of hue. It does so guided by the premise that the visual cortex initiates the mixing of the cardinal L$-$M and S$-$(L+M) pathways and thereby transforms the discrete cone-opponent signals to a continuous representation of chromatic information. Such mixing mechanisms have been implemented by previous combinatorial models of color processing, though through a largely feed-forward approach or at the level of the single neuron.

Our theory bears in mind the mixing mechanism, but reframes the stage-wise combinatorial scheme to one based on the nonlinear population dynamics within the visual cortex. 
Accordingly, we propose a hue-based cortical connectivity, built upon the cortical hue map micro-architecture revealed by recent optical imaging studies of V1. By considering the intracortical network interactions, we have accounted for V1 cells responsive to the gamut of DKL directions without the need to fine-tune the cortical parameters. We do so without restricting to a particular category of V1 neuron, as both single-opponent and double-opponent, and altogether novel types of cells, have been suggested as the primary messengers of chromatic information. 
Rather, we zoom out from the individual neuron's receptive field to model the aggregate, population-level properties and, in particular, the stable representation of hue. We thereby offer that chromatic processing in the visual cortex is, in its essence, a self-organizing system of neuronal-activity pattern formation, capable of encoding chromatic information in the presence of visual stimuli and generating information in their absence.

Further, in assuming modularity for chromatic processing, we have not ruled out a mechanism for joint feature processing. Our choice to focus on the unoriented color cells within the CO blob regions allowed us to parse out the chromatic pathway for an independent analysis, but should not be interpreted as a claim about its functional independence. We leave open the question of the functional and anatomical separation of the various  streams.

Equally unsettled is the question of how much S cone input contributes to the mixing of the cone-opponent channels, with some studies showing a relatively weak S cone input into the neurons of V1, compared to its L and M cone counterparts \citep{cofd, XiaoS}. The variations across these experiments may stem, in part, from differences in optical imaging and electrode penetration techniques, including the particulars of the chromatic stimulus used \citep{cofd, Liu, limits, Salzmann}. On the whole, however, single-cell recordings have identified two main types of color responsive regions: color patches which contain neurons tuned exclusively to stimuli modulating either of the cone-opponent pathways, and patches with neurons exhibiting a \emph{mixed} sensitivity to a combination of the two \citep{Livingston, LandismanElectro, cofd}. Further experiments on the connectivity between these regions, and among the single- and double-opponent color cell populations of which they consist, may point to added micro-architectures for the hue maps, along the lines of the geometric orientation models of \cite{Horizontal} and \cite{Geometric}.

Finally, we emphasize that the mechanism we offer departs from previous combinatorial color models which predict hue sensation at the final stage of processing \citep{Valois, Mehrani2020}, as well as neural-field models that conflate cone- and color-opponency in their interpretations \citep{FaugerasHall, Smirnova, Song}. The emergent hue tuning curves we have characterized are a network property reflective of the \emph{physiological} neuronal responses, and should not be confounded with our \emph{perception} of hue. A photon of wavelength 700 nm striking a retina is no more ``red'' than any other particle --- color is a perceptual phenomenon not yet represented in these first stages of vision. By recognizing that the hue tuning mechanism of the visual cortex is an early stop in the  pathway, we point to the need for further field theory approaches to our understanding of color perception.

\begin{acknowledgments}
The authors acknowledge the fruitful and stimulating discussions with Wim van Drongelen and Graham Smith. This work was supported in part by NSF grant DMS-2052109 and NIH grant R01 NS118606 to PJT.
\end{acknowledgments}

\appendix

\section{Linear Solution}\label{linearmodel}
We assume in the linear case that the net input $h(\theta,t)$ is above threshold throughout the dynamics such that the activity profile is never cut off and $\mathcal{H}\big(h(\theta,t)-T\big)=1$ $\forall$ $\theta$ $\in$ $\{-\pi,\pi\}$. Equation \ref{ratedecomposed} therefore takes the linear form:
\begin{align}
\begin{split}
\label{ratedecomposedlinear}
\tau_{0}\sum \limits_{\mu =-\infty}^{\infty}\frac{dc_{\mu}(t)}{dt}{\hat{e}_{\mu}(\theta)} =&-\sum \limits_{\mu =-\infty}^{\infty}c_{\mu}(t)\hat{e}_{\mu}(\theta) \\
   &+\beta \big(h(\theta,t)-T\big)
\end{split}
\end{align}

Taking the inner product of \ref{ratedecomposedlinear} with $\hat{e}_{\nu}$ on the full domain $\equiv \{-\pi,\pi\}$, we obtain the system of equations for all the coefficients $c_{\nu}$:
\begin{align}
\label{systemofequationslin}
    \tau_{0}\frac{dc_{0}(t)}{dt}&=(2 \pi \beta J_{0}-1)c_{0}(t)-\sqrt{2\pi}\beta T \nonumber \\
    \tau_{0}\frac{dc_{1}(t)}{dt}&=(\pi \beta  J_{1}-1)c_{1}(t)+\sqrt{\tfrac{\pi}{2}}\beta c(l-is)\nonumber \\
    \tau_{0}\frac{dc_{\text{-}1}(t)}{dt}&=(\pi \beta  J_{1}-1)c_{\text{-}1}(t)+\sqrt{\tfrac{\pi}{2}}\beta c(l+is) \nonumber \\
    \tau_{0}\frac{dc_{\nu}(t)}{dt}&=-c_{\nu}(t) \ \forall \ \lvert \nu \rvert >1
\end{align}

We may thus solve for each of the coefficients independently, yielding equations for the evolution of each. Substitution into the activity expansion \ref{decompose} then gives the closed-form solution for the evolution of the activity:

\begin{widetext}
\begin{eqnarray}
\label{linearsol}
    a(\theta,t)=\big\{K_{0}e^{-(1-2\pi \beta J_{0})t/\tau_{0}}-\tfrac{\sqrt{2\pi}\beta T}{1-2\pi \beta J_{0}}\big\}\tfrac{1}{\sqrt{2\pi}}  + \big\{K_{ \text{-}1}e^{-(1-\pi \beta J_{1})t/\tau_{0}}+\tfrac{\sqrt{\tfrac{\pi}{2}}c \beta (l+is) }{1-\pi \beta J_{1}} \big\} \tfrac{1}{\sqrt{2\pi}} e^{-i\theta} \nonumber \\[9pt]
     + \big\{K_{1}e^{-(1-\pi \beta J_{1})t/\tau_{0}}+\tfrac{\sqrt{\tfrac{\pi}{2}}c \beta (l-is) }{1-\pi \beta J_{1}} \big\} \tfrac{1}{\sqrt{2\pi}} e^{i\theta} 
     + \big\{K_{\nu}\tfrac{1}{\sqrt{2\pi}}e^{-t} \big\}  e^{i\nu\theta}\Big\lvert_{\lvert\nu\rvert>1}
\end{eqnarray}
\end{widetext}
where the constants $K_{\nu}$ are determined by the Fourier coefficients $c_{\nu}(0)$ of the initial activity $a(\theta,0)$. 

For $J_{0}< \frac{1}{2 \pi\beta}$ and $J_{1}< \frac{1}{\pi\beta}$, the solution approaches the globally asymptotic stable steady-state tuning curve
\begin{align}
    \label{linearsteadystate}
    a_{\infty}(\theta)=-\frac{\beta T}{1-2\pi \beta J_{0}}+\frac{c \beta \cos{(\theta-\bar{\theta}})}{1-\pi \beta J_{1}}.
\end{align}
We call the corresponding $(J_{0},J_{1})$ parameter space \textbf{the analytical regime}.

\section{Evolution of Peak Angle}\label{secA1}
\label{peakangle}
We first assume that upon receiving a stimulus $\bar{\theta}$ at time $t=0$, the network has a random spontaneous firing rate $a(\theta, 0)$. Using \ref{decompose}, with $c_{0} \in \mathbb{R}$ and $c_{\mu}=c_{\text{-}\mu}^{*}$, we expand the activity profile in terms of the initial values of the corresponding  coefficients $c_{\mu}(0)$:
\begin{align}
\label{decompose2}
    &a(\theta,0)=  \sum \limits_{{\mu}}   c_{\mu}(0){\hat{e}_{\mu}(\theta)}\nonumber \\[1ex]
    \begin{split}
    &=\frac{1}{\sqrt{2\pi}}\Big\{c_{0}(0)+{\sum \limits_{\scriptscriptstyle{\mu\geq1}}}\big(2c_{\text{-}\mu}^{R}(0)\cos(\mu\theta)+2c_{\text{-} \mu}^{I}(0)\sin(\mu\theta)\big) \Big\} \nonumber 
    \end{split} \\[1ex]
    &=\frac{1}{\sqrt{2\pi}}\Big\{c_{0}(0)+\sum \limits_{\scriptscriptstyle{\mu\geq 1}} 2\big[r_{\mu}(0)\cos(\mu\theta-\phi_{\mu}(0))\big]\Big\}
\end{align}
with $\tan(\phi_{\mu})=\frac{c_{\text{-} \mu}^{I}}{c_{\text{-}\mu}^{R}}$ and $r_{\mu}^{2}=(c_{\text{-} \mu}^{I})^{2}+(c_{\text{-} \mu}^{R})^{2}$ such that $\phi_{\mu}(0)$ are completely determined by the initial conditions. Thus, at $t=0$ the activity profile is composed of an infinite sum of cosine functions, each peaked about a corresponding disparate angle $\phi_{\mu}$, and therefore has no discernible peak. To characterize the evolution of the network activity from these initial conditions to its hue tuning profile at $t \rightarrow \infty$, we seek to obtain the steady-state values of $\phi_{\mu}$ and the corresponding tuning curve peak inductively as follows.

Let us first take $\mu=1$. As seen in Fig. \ref{delta1delta2}, we note that $\delta_1(t)$ and $\delta_2(t)$ are symmetric about $\gamma(t)$ such that $\delta_{2}+\gamma=2\pi-(\delta_{1}+\gamma)$. Using this symmetry, we factor out $\cos(\gamma)$ and $\sin(\gamma)$ respectively in the equations for $c_{\text{-} 1}^{R}$ and $c_{\text{-} 1}^{I}$ in \ref{systemofequationseval}:
\begin{align}
\tau_{0}\frac{dc_{\text{-}1}^{R}}{dt}&= -c_{\text{-}1}^{R}+\tfrac{\beta}{\sqrt{2\pi}}F_{1}\cos{\gamma} \nonumber \\  
\tau_{0}\frac{dc_{\text{-}1}^{I}}{dt}&= -c_{\text{-}1}^{I}-\tfrac{\beta}{\sqrt{2\pi}}F_{1}\sin{\gamma} 
\end{align}
with
\begin{equation}
    F_{1}=\tfrac{c_{h}}{2}\big[(\delta_{2}-\delta_{1})+\sin(\delta_{2}-\delta_{1})\big]+2(T-q_{0})\sin(\gamma+\delta_{1})
\end{equation}
and time arguments suppressed. We let $F^{\star}$ and $\gamma^{\star}$ denote the steady-state values of $F$ and $\gamma$ respectively, allowing for the following expressions for the steady-state values of $c_{\text{-}1}^{R}$ and $c_{\text{-}1}^{I}$:
\begin{align}
\label{steadystatec1}
    c_{\text{-}1}^{R^{\star}}&=\tfrac{\beta}{\sqrt{2\pi}}F_{1}^{\star}\cos\gamma^{\star} \nonumber \\
    c_{\text{-}1}^{I^{\star}}&=-\tfrac{\beta}{\sqrt{2\pi}}F_{1}^{\star}\sin\gamma^{\star}.
 \end{align}
Thus,  we have 
\begin{align}
    \tan(\phi_{1}^{\star})&= \frac{c_{\text{-}1}^{I^{\star}}}{ c_{\text{-}1}^{R^{\star}}} =-\tan(\gamma^{\star}).
\end{align}
Similar calculations for the steady-state values of the higher-order coefficients yield the general equations
\begin{align}
\tau_{0}\frac{dc_{\text{-}\mu}^{R}(t)}{dt}&= -c_{\text{-}\mu}^{R}(t)+\tfrac{\beta}{\sqrt{2\pi}}F_{\mu}(t)\cos\big(\gamma(t)\big) \nonumber \\  
\tau_{0}\frac{dc_{\text{-}\mu}^{I}(t)}{dt}&= -c_{\text{-}\mu}^{I}(t)-\tfrac{\beta}{\sqrt{2\pi}}F_{\mu}(t)\sin\big(\gamma(t)\big). 
\end{align}
As before, we note that the evolution of $c_{\mu}(t)$, and therefore of $F_{\mu}(t)$, $\forall$ $\mu \in \mathbb{Z}$ depends only on the first-order coefficients $c_{\lvert \mu \rvert \leq1}(t)$. Therefore, the steady-state values of the higher-order coefficients
\begin{align}
    \label{steadystatemu}
    c_{\text{-}\mu}^{R^{\star}}&=\tfrac{\beta}{\sqrt{2\pi}}F_{\mu}^{\star}\cos(\mu\gamma^{\star}) \nonumber \\
    c_{\text{-}\mu}^{I^{\star}}&=-\tfrac{\beta}{\sqrt{2\pi}}F_{\mu}^{\star}\sin(\mu\gamma^{\star})
\end{align}
and the corresponding $\phi_{\mu}$, i.e, 
\begin{align}
    \tan(\phi_{\mu}^{\star})&= \frac{c_{\text{-}\mu}^{I^{\star}}}{ c_{\text{-}\mu}^{R^{\star}}}=-\tan(\mu\gamma^{\star}),
\end{align}
are fully determined by the solution to \ref{systemofequationseval}.

Substitution of \ref{steadystatemu} into \ref{decompose} then gives: 
\begin{flalign}
\begin{split}
    a_\infty(\theta)&=\frac{1}{\sqrt{2\pi}}c_{0}^{\star} +\frac{\beta}{{\pi}}{\sum \limits_{\scriptscriptstyle{\mu\geq1}}}\big(F_{\mu}^{\star}\cos(\mu\gamma^{\star})\cos(\mu\theta) \\
    & ~~~~~~~~~~~~~~ -F_{\mu}^{\star}\sin(\mu\gamma^{\star})\sin(\mu\theta)\big) \\[2ex]
    &=\frac{1}{\sqrt{2\pi}}c_{0}^{\star}+\frac{\beta }{{\pi}}{\sum \limits_{\scriptscriptstyle{\mu\geq1}}}F_{\mu}^{\star}\cos\big(\mu(\theta+\gamma^{\star})\big), 
\end{split}
\end{flalign}
so that $\theta=-\gamma^{\star}$ represents the peak angle of the steady-state profile $a_{\infty}(\theta)$. 

Further, from \ref{h_v_theta}, we have
\begin{align}
    \tan(\gamma^{\star})=-\frac{q_\text{{I}}^{\star}}{q_\text{{R}}^{\star}} 
    =-\frac{c \sin\bar{\theta}-{\frac{\beta}{\pi}}\lambda_{\text{-} 1}F^{\star}\sin\gamma^{\star}}{c\cos\bar{\theta}+{\frac{\beta}{\pi}}\lambda_{\text{-} 1}F^{\star}\cos\gamma^{\star}}
\end{align}
which requires 
\begin{equation}
\gamma^{\star}=-\bar{\theta}.  
\end{equation}
That is, the steady-state peak $-\gamma^{\star}$ is equivalent to the LGN hue input $\bar{\theta}$.

\section{Linear Stability Analysis}
This section presents the mathematical details for obtaining equation \ref{nonlinearstability}.

Adding a small perturbation $\delta a(\theta,t)$ to the steady-state tuning curve and substituting the resulting network activity
\begin{equation}
\label{theperturbation}
    a(\theta,t)=a_{\infty}(\theta)+\delta a(\theta,t)
\end{equation}
into \ref{WC}, we obtain: 
\begin{widetext}
\begin{align}
\label{A9}
\tau_{0}\frac{d\delta a(\theta,t)}{dt}=-\big(a_{\infty}(\theta)+\delta a(\theta,t)\big)+\beta \big(h_{\infty}(\theta)
+\delta h(\theta,t)-T\big)\mathcal{H}\big(h_{\infty}(\theta)+\delta h(\theta,t)-T\big),
\end{align}
\end{widetext}
where $\delta h(\theta,t)$ is a perturbation to the input due to $\delta a(\theta,t)$. Taylor expanding the right-hand side of \ref{A9} in $h(\theta,t) \equiv  h_{\infty}(\theta)+\delta h(\theta,t)$ about $h(\theta,t)=h_{\infty}(\theta)$ then yields
\begin{flalign}
\label{A10}
\tau_{0}\frac{d\delta a(\theta,t)}{dt}&=-\big(a_{\infty}(\theta)+\delta a(\theta,t)\big) \nonumber \\
&+\beta \Big \{ \big(h_{\infty}(\theta)-T\big)\mathcal{H}\big(h_{\infty}(\theta)-T\big) \nonumber \\
&+ \delta h(\theta,t) \mathcal{H}\big(h_{\infty}(\theta)-T\big)+O\big(\delta h^{2}\big) \Big\}.
\end{flalign}
For small perturbations, the higher-order terms in $\delta h(\theta,t)$ are negligible, and, using $a_{\infty}(\theta)=\beta \big(h_\infty(\theta)-T\big)\mathcal{H}\big(h_\infty(\theta)-T\big)$, we rewrite \ref{A10} as
\begin{align}
\tau_{0}\frac{d\delta a(\theta,t)}{dt} = -\delta a(\theta,t)+ \beta \delta h(\theta,t) \mathcal{H}\big(h_{\infty}(\theta)-T\big).
\end{align}

Next, expanding $\delta a(\theta,t)$ as in \ref{pertubationexpansion}, we obtain
\begin{flalign}
\tau_{0}\sum  \limits_{\mu =-\infty}^{\infty}\frac{dD_{\mu}(t)}{dt}{\hat{e}_{\mu}(\theta)} =-\sum \limits_{\mu=-\infty}^{\infty}D_{\mu}(t)\hat{e}_{\mu}(\theta) \nonumber \\
+\beta \delta h(\theta,t)\mathcal{H}\big(h_{\infty}(\theta)-T\big),
\end{flalign}
wherein we express $\delta h(\theta,t)$ in terms of \ref{subs} to yield:
\begin{widetext}
\begin{align}
\label{A12}
\tau_{0}\sum \limits_{\mu=-\infty}^{\infty}\frac{dD_{\mu}(t)}{dt}{\hat{e}_{\mu}(\theta)}=-\sum \limits_{\mu=-\infty}^{\infty}D_{\mu}(t)\hat{e}_{\mu}(\theta)+\beta\big[\delta q_{0}(t)
+\delta q_{\text{{R}}}(t)\cos(\theta)+\delta q_{\text{{I}}}(t)\sin(\theta)\big]\mathcal{H}\big(h_{\infty}(\theta)-T\big).
\end{align}
\end{widetext}

Finally, taking the inner product of \ref{A12} with $\hat{e}_{\nu}(\theta)$, and reformulating the thresholding nonlinearity in terms of the critical cutoff angles $\delta_{1}$ and $\delta_{2}$ as in section \emph{Evolution of Network Activity}, we arrive at 
\begin{eqnarray}
\tau_{0}\frac{dD_{\nu}(t)}{dt}&=-D_{\nu}(t)+\beta\int_{\delta_{1}^{\star}}^{\delta_{2}^{\star}}\big[\delta q_{0}(t)+\delta q_{\text{{R}}}(t)\cos(\phi)\nonumber \\
&+\delta q_{\text{{I}}}(t)\sin(\phi) \big]\hat{e}^{*}_{\nu}(\phi)d\phi,
\end{eqnarray} 
where $\delta_{1}^{\star}$ and $\delta_{2}^{\star}$ are the steady-state values of the critical cutoff angles.

\bibliography{color-tuning1}

\end{document}